%% file: main.tex
\documentclass[manuscript, nonacm]{acmart}

\usepackage{multirow}
\usepackage{multicol}
\usepackage{arydshln}
\usepackage{multirow}
\usepackage{longtable}
\usepackage{array}
\usepackage{colortbl}
\usepackage{xcolor}
\usepackage{soul}
\usepackage{changepage}
\usepackage{enumitem}
\usepackage{arydshln}
\usepackage[outdir=./]{epstopdf}

\definecolor{VA}{HTML}{ffebd2}
\definecolor{user}{HTML}{FFFFFF}
\definecolor{design}{HTML}{F7F8A7}
\definecolor{purple}{HTML}{d9cae8}

\sethlcolor{design} 
\definecolor{mediumgray}{HTML}{808080}

\newcolumntype{M}{>{\ttfamily\small}p{7.5cm}}
\newcolumntype{S}{>{\ttfamily\small}p{5.5cm}}
\newcolumntype{L}{>{\columncolor{VA}\ttfamily\small}p{7.5cm}}
\newcolumntype{X}{>{\columncolor{VA}\ttfamily\small}p{9.5cm}}
\newcolumntype{D}{>{\columncolor{design}\ttfamily\small}p{6cm}}
\newcolumntype{I}{>{\ttfamily\small}p{9.5cm}}

\newcommand{\myc}[2][mediumgray]{%
  \textcolor{#1}{#2}%
}

\def\eg{\emph{e.g., }} 
\def\ie{\emph{i.e., }}

\AtBeginDocument{%
  \providecommand\BibTeX{{%
    \normalfont B\kern-0.5em{\scshape i\kern-0.25em b}\kern-0.8em\TeX}}}

\begin{document}

\title[Situated Understanding of Errors in Older Adults' Interactions with Voice Assistants]{Situated Understanding of Errors in Older Adults' Interactions with \\Voice Assistants: A Month-Long, In-Home Study}

\author{Amama Mahmood}
\email{amama.mahmood@jhu.edu}

\affiliation{%
  \institution{The Johns Hopkins University}
  \streetaddress{3400 N. Charles St}
  \city{Baltimore}
  \state{Maryland}
  \country{USA}
  \postcode{21218}
}

\author{Junxiang Wang}
\email{jwang334@jhu.edu}

\affiliation{%
  \institution{The Johns Hopkins University}
  \streetaddress{3400 N. Charles St}
  \city{Baltimore}
  \state{Maryland}
  \country{USA}
  \postcode{21218}
}

\author{Chien-Ming Huang}
\email{chienming.huang@jhu.edu}
\affiliation{%
  \institution{The Johns Hopkins University}
  \streetaddress{3400 N. Charles St}
  \city{Baltimore}
  \state{Maryland}
  \country{USA}
  \postcode{21218}
}

\renewcommand{\shortauthors}{Mahmood et al.}
\begin{abstract}

Our work addresses the challenges older adults face with commercial Voice Assistants (VAs), notably in conversation breakdowns and error handling. Traditional methods of collecting user experiences-usage logs and post-hoc interviews-do not fully capture the intricacies of older adults' interactions with VAs, particularly regarding their reactions to errors. To bridge this gap, we equipped 15 older adults' homes with smart speakers integrated with custom audio recorders to collect ``in-the-wild'' audio interaction data for detailed error analysis. Recognizing the conversational limitations of current VAs, our study also explored the capabilities of Large Language Models (LLMs) to handle natural and imperfect text for improving VAs. Midway through our study, we deployed ChatGPT-powered VA to investigate its efficacy for older adults. Our research suggests leveraging vocal and verbal responses combined with LLMs' contextual capabilities for enhanced error prevention and management in VAs, while proposing design considerations to align VA capabilities with older adults' expectations.

\end{abstract}







\begin{teaserfigure}
  \includegraphics[width=\textwidth]{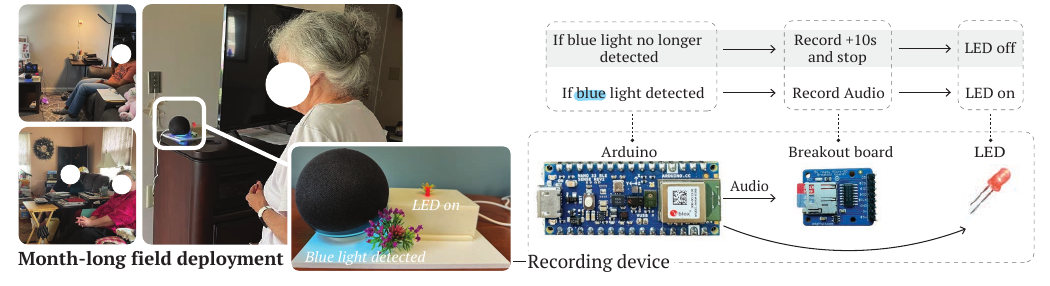}
  \caption{We investigated the interactions of older adults with commercial voice assistants (VAs) in their homes over a period of four weeks with a focus on the emergence of interaction errors and participants' repair strategies. To augment our collected interaction data, we recorded audio using the device depicted above. Each participant's interaction with the VA was recorded in its entirety---plus an additional 10 seconds after the interaction concluded---to capture their spontaneous reactions and responses to the VA. }
  \Description{The figure shows photos of three participants in their homes with the smart speaker deployed with the recording device for a month. There is a blown up picture of the smart speaker and recording device setup that follows a system flow diagram illustrating the functionality of a setup used in a month-long field deployment involving blue light detection and audio recording. At the core of the system is an Arduino wired to both an LED (to indicated that device is recording) and a breakout board for audio recording. The Arduino is placed in proximity to Alexa's blue light to detect when it turns on. The process begins with a decision point: ``If blue light is detected.'' When blue light is detected, the system activates the ``Record Audio'' block and then turns the red ''LED on'' block indicating that audio is being recorded. The flow continues until blue light is no longer detected, at which point the system is programmed to continue recording for an additional 10 seconds before turning off both the recording device and the LED.}
  \label{fig:teaser}
\end{teaserfigure}
\maketitle

\renewcommand{\thefootnote}{}
\footnotetext{This is the author’s version of the work. It is posted here for your personal use. Not for redistribution.}

\renewcommand{\thefootnote}{\arabic{footnote}}







\section{Introduction}

Despite their growing popularity, commercial voice assistants (VAs) are not designed for special populations such as older adults \cite{stigall2019older}, even though they hold the potential to significantly enhance older adults' quality of life and autonomy \cite{smith2023smart, blair2019understanding}.
Data-driven, intelligent VAs are not immune to errors \cite{pearl2016designing}, and older adults' adoption and sustained use of VAs are often impeded by common AI failures \cite{kim2021exploring, pradhan2020use}. Errors are particularly prevalent in their interactions due to older adults' verbose and disfluent speech patterns \cite{mortensen2006age}. Consequently, it is crucial to investigate and analyze the conversational breakdowns specifically experienced by older adults when interacting with these systems.

While researchers have investigated older adults' long-term interactions with VAs at home---focusing on their adoption of the technology, its benefits and challenges, and their perceptions of the experience \cite{kim2021exploring, pradhan2019phantom, pradhan2020use, purao2021use}---a significant research gap remains in the area of comprehensive error analysis.  
A major barrier to studying errors in detail is insufficient error logging in VA usage data---resulting in entries like \textit{``Text not available''} or \textit{``Unknown''} \cite{pradhan2020use}---which often leaves conversational context and real-time user reactions unrecorded. Furthermore, speech recognition errors are often not identifiable solely from transcribed usage logs \cite{kim2021exploring};
supplementary audio recordings are required to capture more nuanced aspects of how smart speakers are embedded into conversational settings \cite{porcheron2018voice}. 
Given that commercial VAs do not provide a mechanism to record complete audio interactions, they fall short in capturing users' immediate thoughts and reactions; such audio data can also be particularly crucial in instances where errors disrupt conversational flow. Documenting how users respond or adapt to continue the conversation can offer insights into not only the nature, frequency, and occurrence of errors, but also users' perceptions of VAs and the strategies they employ to advance the conversation. 
Therefore, we pose the following research question: \textbf{What are the interaction dynamics between older adults and VAs during errors and breakdowns, and how do their interactions evolve over time?}

To overcome the lack of complete interaction data, we designed and developed a custom audio recorder to capture and log user interactions with a commercial VA---Amazon's ``Alexa''---as audio files. We recorded each interaction from start to finish, plus an additional 10 seconds post-interaction to document users' verbal and vocal responses during and immediately after their interactions with Alexa.
We deployed Amazon Echo Dot smart speakers augmented with our recording device at 15 older adults' homes for four weeks to capture rich, ``in-the-wild'' interaction data. 
Our study highlights key issues in older adults' use of VAs, including frequent intent recognition errors that occurred primarily due to the VA's limited understanding of user intent and its failure to accommodate their specific needs (\eg accounting for forgetfulness, preference for a more natural speech style). Furthermore, we found that errors often escalated when older adults attempted to correct them, as these attempts typically failed to address the initial misunderstanding and resulted in numerous additional unresolved errors. We highlight the potential of leveraging users' immediate reactions and responses as implicit cues for VAs to identify errors and improve error management.

Recent advancements in natural language processing have enabled the integration of large language models (LLMs) into voice assistance \cite{alexaLLM2023, OpenAI2023}, significantly improving VAs' capabilities to understand and generate human-like speech thereby reducing erroneous interactions \cite{mahmood2023llm}. This study provided us an opportunity to examine the potential benefits and challenges associated with integrating LLMs into voice assistance, specifically for information retrieval tasks undertaken by older adults within their homes. As such, we integrated ChatGPT into an Alexa skill and deployed it as a technology probe \cite{hutchinson2003technology} in the latter half of our four-week study.
Our preliminary findings emphasize the resilience of LLMs in conversation and illustrate the learning curve necessary for older adults to engage effectively with more sophisticated VAs, providing crucial insights for the design of future LLM-powered VAs. 

Our work's contributions are threefold:

\begin{enumerate}[leftmargin=*]
    
    \item \textbf{Analysis of real-world interaction data:} We qualitatively analyzed ``in-the-wild'' audio interaction data to understand how older adults incorporate VAs into their lives, focusing especially on the nuances of erroneous interactions. Our findings reveal both opportunities and challenges, offering valuable design guidelines for creating more robust and fluid VA interactions. 
    \item \textbf{Data collection tool:} We introduce a novel method for collecting in-the-wild audio data, enabling a deeper understanding of how users interact with and perceive VAs within their personal environments (Fig. \ref{fig:teaser} and \ref{fig:3d-printing}). 
    \item \textbf{LLM-powered VA as a technology probe:} We developed and deployed an LLM-powered VA as a technology probe to explore how older adults engage with conversationally adept, futuristic VAs to identify the opportunities and limitations of LLM integration into VA technologies for this specific population. 
    
\end{enumerate}

\section{Related Work}

Our study explored the integration of voice assistants\footnote{In this manuscript, ``voice assistant'' refers to assistants in smart speakers, ``voice-based'' applies to agents on other platforms smartphones, watches, or web-based systems, and ``voice interface'' refers to the user interaction modality rather than the assistant.} into older adults' daily activities, focusing on errors and the potential impact of LLMs on their information-seeking experience. We review relevant prior work below.

\subsection{VAs in Older Adults' Lives}

VAs have become a staple in many households \cite{Lis2022-ou}, with notably high adoption amongst older adults \cite{Lifestyle2022-yf, Kakulla2021-xj}; VAs not only exhibit significant potential in enhancing their quality of life \cite{smith2023smart, blair2019understanding}, but also have been shown to uplift their spirits, provide more opportunities for mental engagement, and aid in fostering social connections \cite{upadhyay2023studying, oewel2023voice}.
Common uses of VAs among older adults include seeking information, getting weather updates, listening to music, and setting reminders \cite{arnold2022does, o2020voice}. Voice-based agents 
have also been explored in assisting older adults with specific tasks---for example, in interacting with complex user interfaces, where VAs help them locate interface features \cite{yu2023history}, or controlling home appliances \cite{choi2020use, kowalski2019older}---highlighting their preference for voice interfaces over text-based ones \cite{kowalski2019older, wulf2014hands, liu2020design}. 

While initial VA interactions among older adults are often entertainment-oriented, there is a notable shift toward practical applications as they become more familiar with the technology, particularly with regard to health-related queries \cite{pradhan2020use, brewer2022empirical}.
In terms of such task-specific applications, VAs have shown significant promise as health aids; prior research focusing on specific needs such as pain management found VAs to be particularly beneficial for older adults \cite{shade2020voice}, with their abilities to set reminders and promote accountability in health routines especially valuable \cite{shade2020voice, bolanos2020adapting}. Other work has explored health information-seeking \cite{brewer2022empirical, harrington2022s, sanders2019exploring}, preventive care \cite{yamada2018development}, and the potential use of reminders for medication administration \cite{chen2021understanding, bolanos2020adapting, jesus2020voice}.

Apart from VAs' practical use, older adults' social perceptions of these tools are influenced by various factors, such as their familiarity with technology, prior experiences with VAs, and personal preferences. For instance, a user's emotional state and even a smart speaker's physical location can affect whether the user views the assistant as just a machine or as a more humanlike companion \cite{pradhan2019phantom}. As users grow more accustomed to VA technology,
they begin to value not just the systems' operational ease, but also the sense of digital companionship they offer \cite{kim2021exploring}.





\subsection{Conversational Challenges Faced by Older Adults While Interacting with VAs}
\label{sec:background-challenges}
Although older adults recognize the benefits of using VAs, some potential barriers to the adoption of this technology remain for the aging population. For instance, older adults' inability to find utility in smart speakers and their habitual reliance on other ways of completing tasks are mentioned as major factors contributing to their disuse of such devices \cite{trajkova2020alexa}; they also point out potential data protection issues and fear of being patronized and manipulated by VAs  as concerns \cite{horstmann2023alexa}. While VAs have the potential to support older adults' daily activities and well-being, frustrating conversational breakdowns that occur \cite{kim2021exploring} may impact their adoption and continued use of the technology.
Older adults experience higher error rates and often require more time to complete tasks when interacting with voice-based virtual assistants \cite{chiaradia2019evaluating, barros2020usability}, underscoring the increased conversational challenges they face due to factors such as loss of cognitive and motor skills and limited prior technology use \cite{barros2020usability}. 

While research shows that older adults perceive that they have a good mental model of VAs' capabilities (such as the necessity of asking questions in a certain way) \cite{pradhan2020use}, other research points out that they are not fully sure of the extent of those capabilities \cite{kim2021exploring}. Such a mismatch between their perceptions and reality can cause conversational breakdowns, negatively affecting user experience and leading to increased frustration. 
Even though older adults may realize that they need to say things a certain way, they often find it challenging to remember commands associated with VA features and skills \cite{pradhan2020use, stigall2019older}---even those as simple as a wake word \cite{cuadra2022inclusion, angelini2021seniors, pradhan2020use}. Instead, they frequently resort to trial and error following conversational breakdowns \cite{kim2021exploring} and may eventually give up trying to reformulate their queries after multiple failed attempts, leading to incomplete error recovery and thus limiting their use of VAs \cite{pradhan2020use}.

Although they are somewhat cognizant of VAs' limitations---such as their lack of follow-up and contextual understanding---older adults prefer and initially attempt a more conversational interaction style \cite{pradhan2019phantom}. However, frustration arising from VAs' inability to remember context \cite{pradhan2019phantom} often forces older adults to modify their approach. 
This leads to a breakdown, especially in information retrieval (\eg health-related queries), causing older adults to reluctantly shift from their preferred conversational style to more scripted, self-contained queries \cite{brewer2022empirical}. Consequently, many interactions with VAs end up becoming command-based---indicating a transactional, rather than conversational relationship
\cite{purao2021use,purao2019data} and usually lack proactive interchanges \cite{yamada2018development}. These shifts to transactional dynamics have led researchers to debate whether such interactions should truly be considered ``conversations'' in the first place \cite{porcheron2018voice}.

The lack of holistic analysis of daily user-VA interactions (due to the absence of real-world audio interaction data) makes it difficult to concretely interpret the extent of older adults' awareness of VAs' capabilities and their efforts to recover from conversational breakdowns \cite{kim2021exploring}.
Therefore, an in-depth analysis of VA failures in interactions with older adults is necessary to better understand the complexities of their daily interactions. 
While previous research has identified error types, estimated error rates, and examined recovery strategies in older adults' VA use \cite{pradhan2020use, kim2021exploring}, these studies do not address the sequential nature of how errors compound when users attempt to correct them. Our study closes this gap by utilizing recorded interaction data, which captures participants' immediate reactions and actions (\eg recovery attempts) after conversational breakdowns, allowing us to analyze the error ``snowball effect'' and gain a deeper understanding of the dynamics involved in older adults' erroneous interactions with VAs.

\subsection{Research Tools and Methodologies for Longitudinal Field Studies on Older Adults and VAs}

Several rigorous methodologies have been adopted to gather data about older adults' use and perception of VAs in longitudinal at-home studies, which offer
insights into their changing perceptions and interaction dynamics over time. 
One primary approach is the combination of semi-structured interviews \cite{pradhan2019phantom, pradhan2020use, kim2021exploring} and daily diary entries \cite{pradhan2020use}, enabling researchers to gather subjective user perceptions of their experiences with VAs. 
A second approach relies on complementing understanding of user perceptions
with a detailed analysis of usage logs to reveal usage frequency and evolving patterns in VA utilization \cite{purao2021use, pradhan2020use, kim2021exploring, oewel2023voice}. However, this method is constrained by transcription inaccuracies, which complicate error comprehension based solely on usage logs \cite{kim2021exploring}---further suggesting that the usage logs are not always objective, either (\ie they are not always truly reflective of users' interactions with VAs). Moreover, usage logs are unable to capture organic vocal reactions, interruptions, and overlapping speech during VA interactions.
Longitudinal studies on VA technologies for older adults lack real audio interaction data, which has shown to be beneficial for understanding user behavior \cite{porcheron2018voice}; such data can be particularly useful in understanding the nature of conversational breakdowns. Prior work has also highlighted the need for further research on incorporating objective audio data to deepen our understanding of such interactions \cite{purao2021use}. Therefore, in this study, we developed a tool to augment interaction data with real-world audio recordings, thus capturing older adults' verbal and vocal reactions and responses to Amazon's Alexa VA; this feedback also includes interruptions and overlapping speech, which are not logged in traditional usage data.

\subsection{Integration of LLMs into Voice Assistance}
Despite older adults' desires to have conversational interactions with VAs, current commercial VA capabilities are far more limited \cite{pradhan2019phantom}.
LLMs, on the other hand, have shown enhanced conversational capabilities, yielding robust and fluid interactions with users \cite{mahmood2023llm, chan2023mango}.
Recent efforts to incorporate LLMs into voice-based interactions for older adults have shown improvement in conversational aspects \cite{yang2024talk2care, jo2023understanding}; LLMs have also shown potential in absorbing various speech recognition errors \cite{mahmood2023llm}. In this work, we explore if incorporating LLMs into VAs can reduce error rates for older adults.  



\section{Methods}
To understand older adults' perceptions and interactions with VAs, we conducted a four-week field deployment of a smart speaker, the $5^{th}$ generation Amazon Echo Dot. We collected data via usage logs, audio interactions from a novel recording device, and semi-structured interviews at the beginning and end of deployment. This section outlines the implementation of our recording device and  integration of ChatGPT into an Alexa skill, followed by a description of our longitudinal deployment, participant details, and data analysis approach.

\subsection{Data Collection: Recording Device}

In our study, we utilized a specially developed recording device to capture users' interactions with Alexa, including their immediate responses and reactions 
following each voice interaction. The recording device comprises of three modules: 

\begin{enumerate}[leftmargin=*]
    \item \textbf{Detector:} The recording device detects the activation of Alexa---indicated by the blue light ring on the bottom of Amazon Echo Dot---via the light color detector on an Arduino NANO BLE SENSE 33. 
    
    \item \textbf{Recorder:} Triggered by the detection of the Echo Dot's blue light, the Arduino begins recording audio onto a microSD card inserted into a connected Adafruit breakout board. Recording ceases 10 seconds after the blue light turns off. However, if the user reactivates Alexa within this 10-second window, the device continues recording until the end of the new interaction, plus an additional 10-second period.

    \item \textbf{Privacy indicator:} To address privacy concerns and enhance user comfort, our device does not continuously buffer audio for keyword detection, as utilized in 
    prior research \cite{porcheron2018voice}. Instead, it only buffers audio when the blue light is detected at the start of an interaction. To ensure transparency, a red LED light positioned on top of the recording device illuminates to indicate active audio recording, allowing the user to closely monitor recording status. 

\end{enumerate}

\begin{figure*}[t]
     \includegraphics[width=\textwidth]{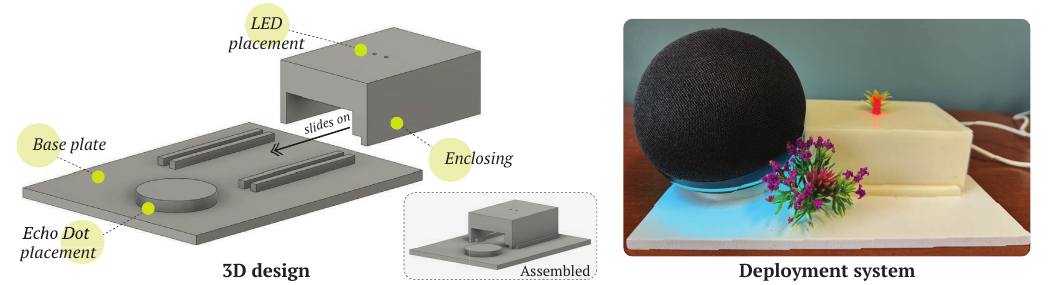}    
     \caption{Design of the enclosure and base plate for placing the Echo Dot smart speaker and recording device. CAD files are available online at \url{https://bit.ly/3UIgAri} for 3D printing.}
    \Description{This figure illustrates the design and assembly process for a 3D-printed deployment system that involves a base plate, a rectangular box enclosing to place Arduino audio recorder in it, a cylindrical placement for an Echo Dot device, and holes for LED indicator on the top of the rectangular box enclosing. The figure shows how the enclosing slides on the base plate that has placeholders for the enclosing. The echo dot placement is right next to the enclosing so that the Arduino can capture the blue light. In the middle of the figure the assembled 3D model is shown. Towards the right of the figure is the actual photo of the deployment system with echo dot placed on the printed 3D structure and the LED light is visible on the top of the enclosing. The blue light and the LED light are on showing that the smart speaker is activated and the device is recording audio. There is a decorative small flower bush right in front of where echo dot touches the enclosing.}
    \label{fig:3d-printing}
\end{figure*}

The recording device is adaptable for use with other smart speakers, as integration with wake word detection is also feasible. We also designed and 3D-printed a compact enclosure for the recording device, which can be mounted on a base plate with a designated spot for the Echo Dot, as shown in Fig. \ref{fig:3d-printing}. The code for the recording device, 3D printing sketch, and assembly instructions will be publicly available\footnote{Link to the code and design for our recording device and additional study materials: \url{https://bit.ly/3UIgAri}.}.
To validate the functionality of our recording device, we conducted a pilot test with one young adult and one older adult by deploying a smart speaker and our recording device in their homes for one week. This test confirmed that the device operated as intended. 

\subsection{Technology Probe: Integrating ChatGPT into an Alexa Skill}

To investigate older adults' interactions with more conversational VAs and to examine the associated benefits and challenges thereof, we designed an LLM-powered VA by incorporating ChatGPT-3.5 into an Amazon Alexa skill, similar to prior work \cite{mahmood2023llm}. We refer to this skill as the ``ChatGPT+Alexa skill'' or simply ``ChatGPT skill.''
This implementation, designed for general information retrieval and deployed as a technology probe, aims to reveal the potential advantages and limitations of LLM-powered VAs. 
Adopting the technology probe framework in \cite{hutchinson2003technology}, we focused on investigating the usability and effectiveness of an LLM-powered VA in real-world settings. 

\subsection{Longitudinal Study: Procedure}
The study comprised a four-week field deployment that included an initial interview, device setup, a demo session, a check-in at Week 1, the deployment of our LLM-powered Alexa skill at Week 3, and a final session at Week 4 involving a semi-structured exit interview. 

\subsubsection{Week 1: Initial interview and smart speaker setup.} 
Taking place at the beginning of Week 1, the first session consisted of following activities in given order:

\begin{enumerate} [leftmargin=*]
    \item \textit{Introduction and consent.} The experimenter explained the study and obtained consent. 
    \item \textit{Demographics survey.} Participants completed a demographics survey
    (Table \ref{tab:participants}) and a short quality of life survey 
    \cite{bowling2013short}.
    \item \textit{Pre-study interview.} The experimenter conducted a brief semi-structured interview to gauge participants' awareness and perception of smart technologies in general and of VAs (\eg Amazon's Alexa and Apple's Siri).
    \item \textit{Smart speaker setup.} The experimenter set up a smart speaker at each participant's desired location. 
    The experimenter informed participants that they could move the speaker and demonstrated how. An instruction page for debugging internet issues was provided. Participants could additionally contact the experimenter at any time via email, text, or phone call. 
    At this time, the Amazon Alexa app was installed in their device of choice (\eg smart phone or tablet); 
    the experimenter set up participants' name pronunciations, voice profiles, and device location, and turned on the built-in adaptive listening feature 
    and follow-up mode in the smart speaker's settings. (The adaptive listening feature provides users an extended duration with which to complete their query before Alexa initiates its response; follow-up mode allows them to ask Alexa follow-up questions without repeating the wake word.) A free Amazon Music subscription was also made available for all participants.
    \item \textit{Demonstration of Alexa and recording device.} The experimenter gave participants an introductory tutorial that covered fundamental VA capabilities including tasks like setting alarms, reminders, and timers; creating shopping and to-do lists; playing music; asking to hear jokes; posing questions; and engaging in casual, unstructured question-and-answer sessions. Furthermore, the experimenter demonstrated the process of accessing and utilizing various Alexa skills. A printed copy of this introductory tutorial was left with each participant\footnotemark[2].
    While showcasing Alexa's capabilities, the experimenter also explained that the audio recorder only records while the red LED light is on. During the demonstration, the experimenter also removed the microSD card from the recording device and played a few recordings for the participants to confirm they correctly comprehended what audio would be recorded during their interactions with the smart speaker.

\end{enumerate}

At the end of Week 1, the experimenter reached out to participants over the phone to inquire briefly about their smart speaker use thus far, mainly to see if the participants had experienced any technical difficulties. 

\subsubsection{Week 3: ChatGPT deployment.}

At the beginning of Week 3, we remotely deployed the implemented ChatGPT skill. Comprehensive instructions, along with illustrative examples and suggested conversation prompts, were provided to participants\footnotemark[2]; these instructions were communicated either via email or in print through our liaison at the community center. The experimenter then contacted all participants over the phone to introduce and demonstrate the skill as a means to hold more natural conversations with a VA that is capable of responding to follow-up questions and statements. This skill was available for them throughout week 3 and 4 of the study. 


\begin{table}[tb]
\centering
\caption{Demographics for our field study. All participants were fluent in English. All were retired except P10a (employed full-time) and P11a (employed part-time). Overall quality of life (QOL) is self-reported; their response to one question. }
\label{tab:participants}
\begin{tabular}{lllllllll}

\textbf{P} &\textbf{Gender} & \textbf{Age} & \textbf{Ethnicity} & \textbf{Disabilities/Aids} & \textbf{ Overall QOL} & \textbf{Prior VA Use} & \textbf{VA Placement}\\
\midrule[1.5pt]
\multicolumn{8}{l}{\textbf{Community Center -- Assisted Living (by self)}} \\
\hline
2   & M & 66   & Prefer not to say  & Wheelchair  &   Alright & Never & Studio\\

3   & F & 82   & Caucasian & Hearing aid &   Alright & Never & Studio\\

4   & M & 79   & Caucasian &  Wheelchair &    Good & In the past &  Studio\\
\hline
\multicolumn{8}{l}{\textbf{Community Center -- Independent Living (by self)}} \\
\hline
5   & F & 77   & Caucasian &  & Good & Never & Living room\\

6   & F & 81   & Caucasian & Cane/Rollator & Alright & Never & Living room\\

7   & F & 74   & African American &   & &  Never & Living room  \\

14  & F & 73   & African American & &  Good & Siri  & Living room\\

15  & F & 73   & African American &   &  Good & Never & Living room\\
\hline
\multicolumn{8}{l}{\textbf{Homeowner (by self)}} \\
\hline
8   & F & 84   & Caucasian & Walker & Alright & Never & Bedroom \\

9   & M & 75   & Caucasian & &   Very good & Alexa & Dining room\\

13  & M & 81   & Caucasian &   Vision issues  &  Good & Never & Dining room  \\
\hline
\multicolumn{8}{l}{\textbf{Homeowner (Couple, with spouse)}} \\
\hline
1a & M & 76   & Caucasian  &   & Good & Alexa, multiple & Living room \\

1b & F & 75   & Caucasian & &   &  Alexa; multiple & Living room \\
\cdashline{1-8}
10a & M & 68   & Caucasian &   &  Good & Never & Living room\\

10b & F & 94   & Caucasian& Walker/scooter &   Bad & Never & Living room\\
\cdashline{1-8}
11a & M & 72   & Caucasian &  &   Very good & Siri  & Kitchen \\

11b & F & 71   & Asian & &    Very good & Siri & Kitchen \\
\cdashline{1-8}
12a & M & 75   & Caucasian  & &  Very good &  Alexa; multiple & Living room \\

12b & F & 72   & Caucasian  & &  Good &  Alexa; multiple & Living room  \\
\cdashline{1-8}

\hline
\end{tabular}
\end{table}
\subsubsection{Week 4: Collection of smart speaker.}

At the end of Week 4, the study concluded with smart speaker collection---unless the participant chose to keep theirs as compensation---and a semi-structured exit interview aimed at gathering insights from the participants regarding their four-week experience with the smart speaker. Participants were encouraged to provide comprehensive information about their overall experience, the VA's usefulness in their daily activities, specific tasks that were facilitated by the VA, features they found enjoyable or useful, and any challenges they encountered.

\subsection{Participants}

We deployed 15 smart speakers in three waves for four weeks each.
Recruitment was based from a community center attached to both independent and assisted living communities; 
we also recruited community-dwelling adults from previous connections. Table \ref{tab:participants}  summarizes participants' demographics and awareness of VA technology. Participants were compensated with either the option to keep their smart speaker or an amount equivalent to the market price of the Amazon Echo Dot (\ie \$50). 
All participants except three---couple P11,  
P5, and couple P10
---kept the smart speaker. This study was approved by our institutional review board. 

\subsection{Data and Analysis}

Audio recordings from each device were transcribed and matched to usage logs from the Amazon Alexa dashboard. We also noted non-verbal and vocal cues such as laughter or filler noises (\eg ``huh,'' ``hmm'') in our transcripts. We took note of any interruptions, overlapping conversations, and comments or remarks about interactions---especially those occurring around interaction errors.
We collected a total of 20 hours and 40 minutes of usage logs and audio recordings from the devices; audio data for participant P6 was missing due to a recorder malfunction. 
For the audio data and Alexa usage logs, we identified 2552 user query-VA response pairs (also referred to as ``turns'') across all participants.

Initially, the first author examined the interaction data, using an inductive approach to identify preliminary codes to gain a broader understanding of conversational breakdowns. 
Following this, two researchers independently coded the data for two participants (approximately 10\% of the total) using these initial codes. They then convened to discuss and resolve any conflicts, resulting in the formulation of the final codebook given in  Appendix \ref{app:codebook}. All transcripts were subsequently analyzed with the established codes deductively applied to each user query-VA response pair.
Each query-response pair was coded for its purpose (type of interaction as informed by prior work \cite{pradhan2020use, kim2021exploring}), the presence and type of error (as defined in Table \ref{tab:errors-def}), whether the error was resolved, and the user's recovery method (\eg moving on, repetition, or clarification). We also noted whether an error led to a conversational breakdown and if it was evident that the participant identified the error from their reaction(s) or actions (by initiating recovery).    
All participant interviews were transcribed and participants' direct quotes were used 
to support, explain, and contextualize our findings. 


\begin{table}[tb]
    \centering
    \caption{Errors and their definitions, grouped by category, along with their occurrences in the interaction data and resolution rate in the next immediate user retry. $^*$Percentage of time participants reacted to each error type.}
    \label{tab:errors-def}
    \begin{tabular}{lp{9cm}lll}
\toprule
Error &  Definition &  Count    & Resolved & Reactions$^*$ \\
\midrule
\midrule
\textbf{Human errors}&\textbf{User made a mistake } & \textbf{72 }& \textbf{38.9\%} & \textbf{6.9\%} \\
\cdashline{1-5}
Wrong wake word      & User used the wrong wake word & 49 & 30.6\% & 8.2\% \\
Partial query       & User did not complete their query & 23 & 56.5\% & 4.3\% \\
\midrule
\textbf{Speech errors} & \textbf{Errors caused due to speech recognition inaccuracies} & \textbf{154} & \textbf{31.8\%} & \textbf{15.6\%} \\
\cdashline{1-5}
Not listened        & User query was not listened by Alexa & 69 & 36.2\% & 4.3\% \\
Mis-trigger             & Alexa was triggered when user did not intend as such& 15 & 6.7\% & 40\% \\
Partially listened    & Alexa captured user query partially & 29 & 27.6\% & 20.7\% \\
Interruption        & Alexa interrupted the user query & 1 & 100\% & 100\%\\
Transcription       & User query was transcribed inaccurately & 40 & 35\% & 20\% \\
\midrule
\textbf{VA errors} & \textbf{Errors attributed to the processing of accurately captured speech} & \textbf{406} & \textbf{20.44\%} & \textbf{14.3\%} \\
\cdashline{1-5}
System              & User query was captured accurately but Alexa failed for an unknown reason and responded with a default phrase such as \textit{``I'm having trouble getting that right now. Please try again.''} & 15 & 26.7\% & 13.3\% \\
Not handled         & User query was captured accurately but Alexa did not respond at all& 57 & 17.5\% & 0.0\% \\
Limitation          & User query was accurately captured but Alexa failed to fulfill the request, responding with a default phrase,  \textit{``Sorry, I don't know about that.''}& 130 & 10.8\% & 14.6\% \\
Intent recognition             & User query was captured accurately but Alexa failed to recognize user intent and responded inaccurately & 204 & 27\% & 18.1\% \\
\bottomrule
\end{tabular}

\end{table}

\section{Findings}

Our dataset has a total of 2,552 one-turn (user query-VA response pair) interactions;  
the interaction data highlights the varying nature of participant engagement with the VA, ranging from brief, singular commands to more extended dialogues (details in supplementary materials).  
Overall, we observed a variety of interactions---from setting reminders to seeking information about various topics---as shown in Tables \ref{tab:interaction_categories} and \ref{tab:question-type}. Since the focus of this paper is on conversational errors, our analysis on participants' usage patterns and the social nature of their interactions captured with our recording device are provided in supplementary materials. 
In this section, we present 
a detailed analysis of participants' erroneous interactions with the VA, followed by additional analysis of their interactions with our LLM-powered VA.








\subsection{Analysis of Erroneous Interactions}

\label{sec:errors}
\begin{figure*}[t]
     \includegraphics[width=\textwidth]{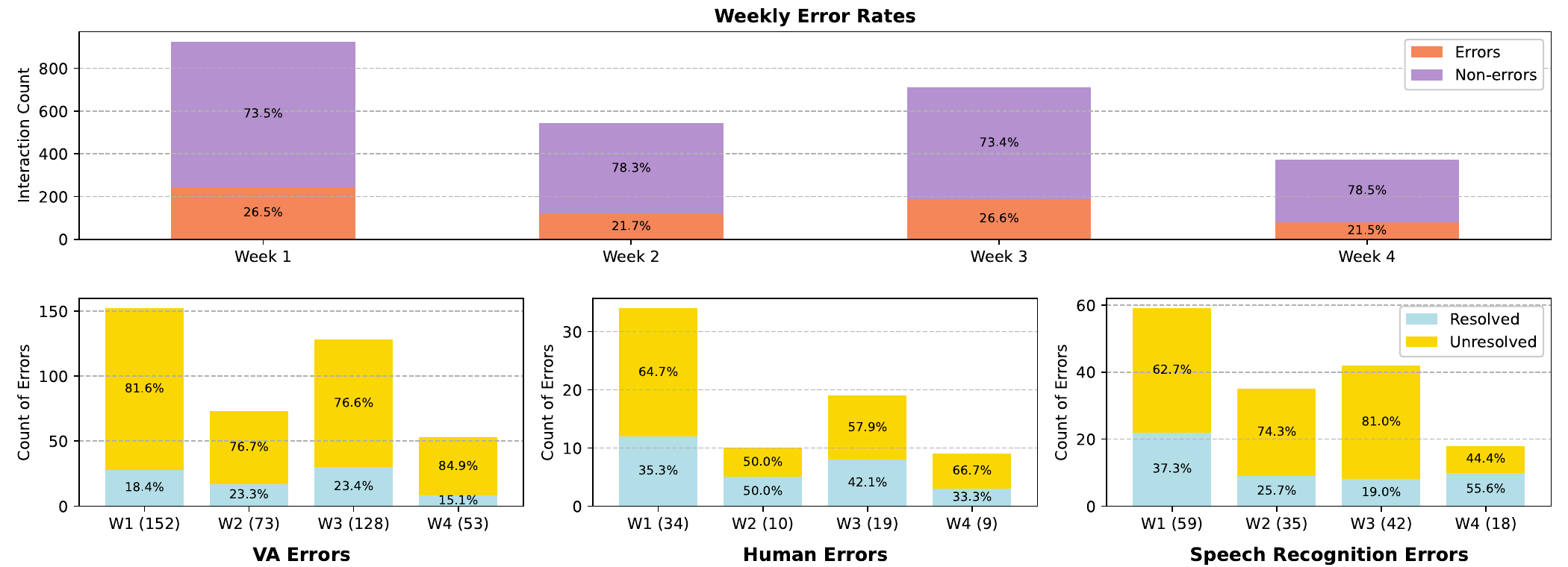}    
     \caption{Error categories and their resolution upon immediate participant retry attempt across four weeks.}
     \Description{The figure contains four stacked bar plots. One in the top row and three in the bottom row. The horizontal axis of each plot represents the progression of time (in weeks) as Week 1, Week 2, Week 3 and Week 4. For the top graph, the vertical axis is labelled as interaction count and the height of bar graph shows total number of interaction in respective week. The bar plot is further stacked to show the number of errors and non-errors in those interactions. The stacked bars are labelled with percentage of errors and non-errors. Throughout the four week the total number of interactions change; Week 1 has highest close to 900 decreasing to Week 4 that lowest less than 400 but Week 3 had more than week 2. Despite that change, the percentage of erroneous interactions is consistently between the range of 21-26 percent. 
     The bottom three plots are for specific error categories: first one VA errors, second one human errors, and third one speech recognition errors. The vertical axis is labelled as count of errors in these plot. The height of the bars show the total count of errors in respective week. The bar plot is further stacked to show the errors that were resolved on first retry vs. those that were not. The stacked bars are labelled with percentage of errors resolved and not resolved. Throughout the three bar plots for error categories, even thought the number of errors change (follow similar trend to the top graph) but the percentage of errors resolved across weeks is consistent. Human errors were resolved the most while VA errors had lowest resolution rates. }
    \label{fig:error-resolution-weeks}
\end{figure*}

We conducted a detailed analysis of errors and users' experience of and behavior surrounding them---an area less explored in prior research. By examining users' immediate responses and efforts to rectify errors as captured via our recording device, we were able to explore the intricate dynamics of user interactions during conversational breakdowns. \\
\textbf{True error rate in VA interactions.} Our study revealed that a considerable 24.76\% of user-VA one-turn queries ($n=632$ out of 2552) had errors, equating to almost one in every four queries failing. Of these, 98.10\% manifested as conversational breakdowns.
This number represents the true error rate, contrasting with prior work \cite{kim2021exploring} where activation errors (among others) were unaccounted for, as they typically are not captured in user logs. However, our methodology, which included the use of a recording device, enabled us to capture these activation errors (\eg wrong wake word, mis-trigger, not listened, partially listened, etc.), providing a more comprehensive understanding of the true error rate in VA interactions. 
While coding the data, we noted whether an error was resolved in an immediate retry; only $25.47\%$ ($n=160$) of all errors were resolved in the immediate next attempt by participants, indicating that many errors either remained unresolved or required multiple attempts to rectify. \\
\textbf{Consistent errors rates over time.} We observed that error rate remained consistent, between 20--30\%, across four weeks despite a varying number of interactions (Fig. \ref{fig:error-resolution-weeks}, top). 
However, error rates in Weeks 1 and 3 were slightly higher than in Weeks 2 and 4, possibly due to the novelty effect from the introduction of the VA at the beginning of Week 1 and the novelty and Hawthorne effects from the introduction of a new feature (\ie the ChatGPT+Alexa skill) at the beginning of Week 3, which resulted in exploratory user behavior aimed at testing the boundaries of the system. 

Below, we report our findings on types of errors and their resolution, the distribution of errors across interaction types, the indication of error recognition by participants through their actions and reactions, observed recovery strategies, and the compounding nature of errors. We also demonstrate how various aspects of interaction errors evolved over time.
\begin{table}[tb]
\centering
\caption{Interaction categories and types with their respective counts and success rates.}
\label{tab:interaction_categories}
\begin{tabular}{llrr|llrr}
\toprule
\textbf{Category} & \textbf{Type} & \textbf{Count} & \textbf{Success} & \textbf{Category} & \textbf{Type} & \textbf{Count} & \textbf{Success} \\
\midrule
\midrule
\multirow{8}{*}{Functional} 
 & Reminder & 251 & 48.60\% &\multirow{9}{*}{Entertainment} & Music & 324 & 79.01\% \\
 & Weather & 205 & 83.41\% &  & Radio & 232 & 73.70\% \\
 & Timer & 174 & 90.23\% &  & Alexa skills & 89 & 58.43\% \\
 & Alarm & 98 & 80.61\% & & Joke & 22 & 86.36\% \\
 & Command & 94 & 94.68\% &  & TV & 10 & 20.00\%\\
 & List & 90 & 77.78\% &  & Story & 7 & 71.43\%\\
 & Communication & 71 & 98.59\% & & Poem & 3 & 33.33\% \\
 \cdashline{6-8}
 & Alexa capabilities & 69 & 52.17\% & & \textbf{Total} & \textbf{753} & \textbf{71.84\%}\\
 \cline{5-8}
 & Time & 46 & 100.00\%  & Information-seeking & Question & 414 & 57.25\% \\
\cline{5-8}
 & Repeat & 16 & 62.50\% & Social & Greetings  & 174 & 93.68\% \\
 & Calendar & 12 & 58.33\% & & \& gratitude & & \\
 \cline{5-8}
 & Notes & 1 & 100.00\% & ChatGPT & ChatGPT & 64 & 68.75\% \\
 \cdashline{2-4} \cline{5-8}
 & \textbf{Total} & \textbf{1130} & \textbf{75.93\%} & Other & Misc. & 17 & 0.00\% \\

\bottomrule
\end{tabular}
\end{table}

\begin{table}[tb]
    \centering
    \caption{Type of information-seeking questions asked. n = count; ER = error rate (percentage of queries of a particular question type that resulted in an error). For instance, $59.62\%$ of total ``local information'' questions resulted in at least one error ($n=31$ out of 52).}
    \label{tab:question-type}
    \begin{tabular}{lrr|lrr|lrr}
        \toprule
 \textbf{Type of question} & \textbf{n} & \textbf{ER} & \textbf{Type of question} & \textbf{n} & \textbf{ER} &\textbf{Type of question}& \textbf{n} & \textbf{ER}\\ 
        \midrule
        \midrule
        Entertainment        & 98  & 28.57\% &Miscellaneous              & 38 & 44.74\% &  STEM-related         & 16 & 31.25\% \\ 
        Local information    & 52  &  59.62\% & Food \&  drink       & 37 & 21.62\%& Specific places      & 14  & 64.29\% \\ 
        Meaning \& definitions   & 45  &  42.68\% & Famous personalities & 31 & 58.06\%  & History              & 10 & 60.00\% \\ 
        Health               & 43  & 55.81\% &  Alexa-related        & 21 & 52.38\% & Travel               & 9  & 11.11\% \\ 
        \bottomrule
    \end{tabular}
\end{table}


\subsubsection{Types of errors and their resolution rates}
\label{sec:error-types-resolution}
Our interaction data encompassed a diverse range of errors (Table \ref{tab:errors-def}), which we broadly categorized into VA, speech recognition, and human errors: \\
\textbf{VA errors.} 
We found that the most frequent errors encountered by participants while interacting with Alexa were intent recognition errors, accounting for $32.3\%$ ($n=204$) of all 632 errors (Table \ref{tab:errors-def}); although Alexa accurately transcribed the user's query in these cases, the response or action provided did not align with the participant's original intent. The next most common error category involved Alexa's limitations in handling user requests ($n=130, 20.6\%$). 
VA errors, while being most frequent, had the lowest resolution rates ($20.44\%$) as opposed to speech recognition and human errors (Table \ref{tab:errors-def}), further highlighting Alexa's limited capabilities in correcting errors despite its attempts at recovery.\\
\textbf{Human error.} On 23 occasions ($3.6\%$), an error occurred because a participant did not complete their query. Participants also incorrectly used ``Alexis'' instead of ``Alexa'' ($n=49, 7.8\%$), a common confusion noted previously \cite{pradhan2020use, cuadra2022inclusion, angelini2021seniors}; the system was able to pick up the wrong wake word as a potential trigger (thus, picked up by our recording device), but marked it as \textit{``Not intended for Alexa.''} Human errors were the least frequent and had the highest resolution rate ($38.9\%$). \\
\textbf{Speech recognition errors.}
Speech recognition errors such as not listened ($n=69, 10.9\%$), transcription errors ($n=40, 6.3\%$), and partially listened queries ($n=29, 4.6\%$) 
were less frequent. Moreover, we observed that resolution success was higher for such errors ($31.8\%$) as opposed to VA errors, suggesting that the primary challenges of smart speaker use revolve around understanding user intent and taking requisite action rather than fixing listening inaccuracies.\\
\textbf{Consistent over time.} We noticed that there was no clear pattern of improvement in resolution rates over the span of four weeks across all error categories (Fig. \ref{fig:error-resolution-weeks}, bottom); the lack of apparent improvement in error resolution indicates that participants struggled with errors regardless of gaining familiarity with voice assistance through direct experience.

\begin{figure*}[t]
     \includegraphics[width=\textwidth]{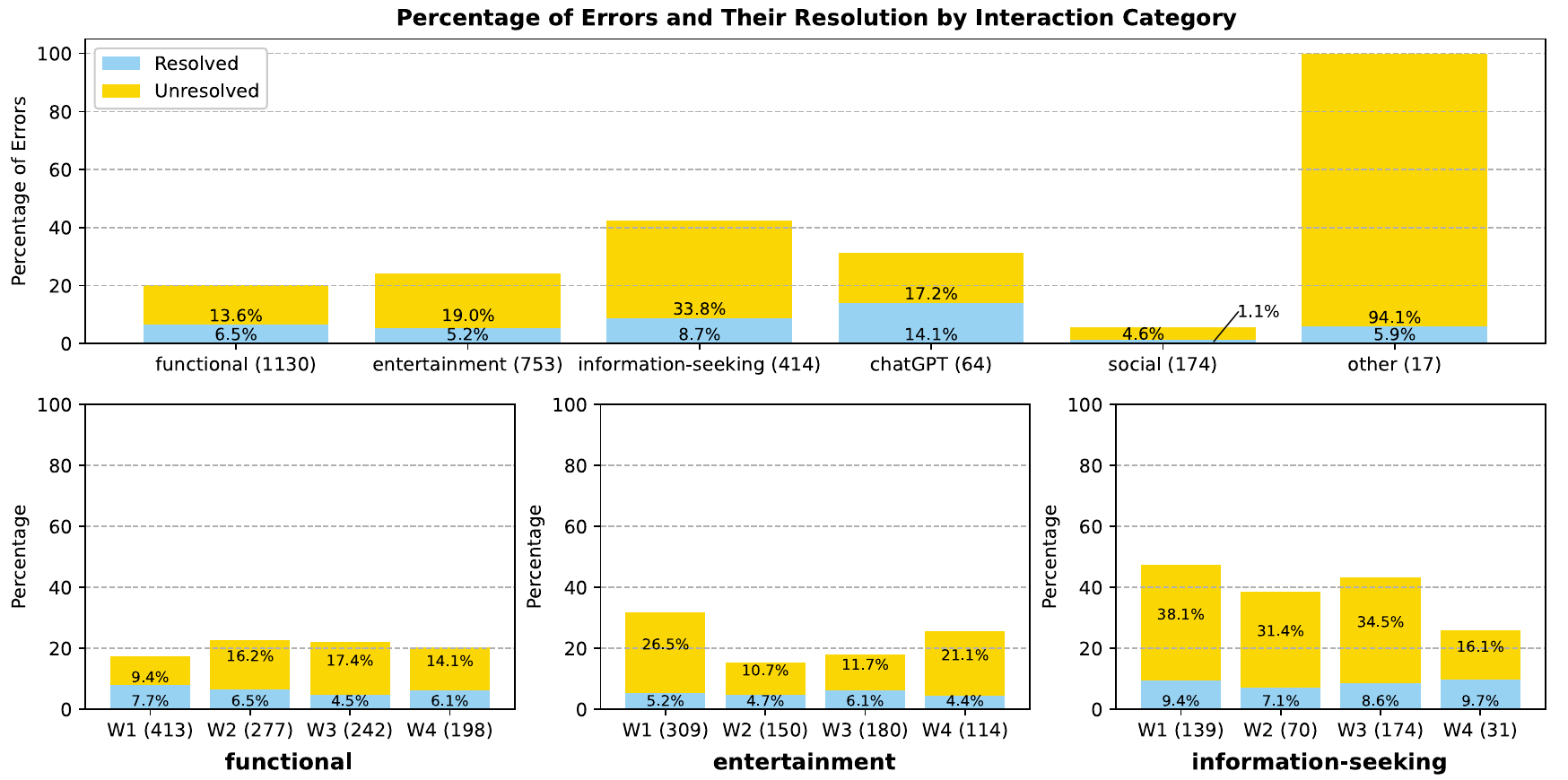}    
     \caption{Error resolution across interaction types over time. The top plot shows for each interaction category the percentage of interactions that resulted in errors and the percentage of interactions in which the errors were resolved; for instance, 414  interactions were categorized as ``information-seeking'' ($100\%$), $\sim44\%  (\sim34\% + \sim9\%)$ of those interactions resulted in an error, and only $\sim9\% $ of the original 414 were resolved after the first retry. The remaining plots for the top three individual categories show how their error resolution rates changed over the full four weeks. 
     Note: Not all errors had retries and some had multiple retries, as discussed later.}
     \Description{The figure contains four stacked bar plots. One in the top row and three in the bottom row. For the top plot, the horizontal axis represents the interaction categories with total number of interactions in bracket with it: functional (1130) entertainment (753), information-seeking (414), chatGPT (64), social (174) and other (17). The vertical axis is labelled as the percentage of errors. The bar plot is stacked to show percentage of resolved and unresolved errors where 100\% represents all interactions. The stacked bars are labelled with percentage as well. The plot shows variable percentage of errors and resolution for all categories which other having highest error rate near 100\% while information seeking has second highest around 44\% and social having least around 6\%. The ChatGPT has highest resolution rate if 14.1\%. 
     The bottom three plots are for top three most frequent interactions categories: first one functional, second one entertainment, and third one information-seeking. For the bottom three plots, the horizontal axis of each plot represents the progression of time (in weeks) as Week 1, Week 2, Week 3 and Week 4. The vertical axis is labelled as percentage of errors. The bar plot is stacked to show percentage of resolved and unresolved errors where 100\% represents all interactions. The stacked bars are labelled with percentage as well. The three plots show that there is no distinct pattern of reduction or increase in error rate or resolution rate within each category over four weeks. 
     Error rate stayed around 20\% for functional and resolution rate between 4-7\%. Error rate stayed around 20\% for entertainment and resolution rate between 4-6\%. Error rate for information seeking remained around 40\% and resolution between 7-10\%. }
    \label{fig:error-interaction-category-weeks}
\end{figure*}

\begin{table}[tb]
\centering
\caption{Example conversations related to errors. `` \dots'' represents the truncation of a conversation for brevity.}
\label{tab:alexa-dialogue-errors}
\begin{tabular}{p{0.3cm} M L}
\textbf{Tag} & \textbf{Participant} & \textbf{Alexa} \\
\midrule
\midrule
C1 & What are Joe Biden's roots? & According to an Alexa Answers contributor: Ireland.\\
\cdashline{2-3}
P2 & Alexa, that's his lineage. I wanna know about his mother and his father. & \myc{[Not listened by Alexa]} \\
 \cdashline{2-3}
 & Alexa, tell me about virus family, his parents. \myc{[Transcription error; he actually said: ``Alexa, tell me about Joe Biden's family, his parents.''] }& From cdc.gov: Mayaro virus (MAYV), the cause of Mayaro fever, is a member of the genus Alphavirus, family Togaviridae \dots \\
 \cdashline{2-3}
 & Alright Alexa, you did a really good job at answering a question I did not ask you. & \myc{[Not listened by Alexa]} \\
 \cdashline{2-3}
 & Alexa, tell me about Joe Biden's family, his mother, his father. & Joe Biden's children are Ashley Biden, Beau Biden, Hunter Biden, and Naomi Biden. His parent is Jean Biden, his siblings are Francis Biden \dots\\
 \cdashline{2-3}
 & Alexa, show his mother was a single mother. \myc{[Transcription error; he actually said: ``Alexa, so his mother is a single mother?'']} & From educalingo.com: The definition of ``single mother'' in the dictionary: A mother who has a dependent child or dependent children and who is widowed \dots\\
 \cdashline{2-3}
 & Alexa, Joe Biden's family was---did he have a father? & Joe Biden's father is Joseph R. Biden, Sr. \\
 \cdashline{2-3}
 & What did he do? & \myc{[Not listened by Alexa]} \\
 \cdashline{2-3}
 & Alexa, what was Joe Biden, Sr.'s educational background and what did he do? & This might answer your question: Joe Biden was educated at Syracuse University and the University of Delaware. \\
 \cdashline{2-3}
 & Alexa, what was Joe Biden's father's background? & Hmm, I don't know that one. \\
 \cdashline{2-3}
 & Alexa, did Joe Biden know his father? & Joe Biden's father is Joseph R. Biden, Sr. \\
 \cdashline{2-3}
 & What what was Joseph F. Biden's educational background? & Before his political career, former Vice President Joe Biden was an attorney. In 1965, he received \dots \\ 
 \cdashline{2-3}
 & Alexa, stop. I'm talking about Joe Biden's father. What was his educational background? & Sorry, I don't know that one. \\
 \cdashline{2-3}
\hline
C2 \newline P2 & Alexa, do you have a memory of what I say to you? & I didn't find any stickies. If you'd like to create one, just ask me to make a new sticky note.\\ 
\cdashline{2-3}
 & Alexa, I don't want a sticky note, I want you to remember the conversation. Do you have any kind of memory about what I asked you? & Okay, I’ve saved your sticky. \myc{[Intent error]} \\
 \cdashline{2-3}
 & Alexa, never mind. & \myc{[No response from Alexa]} \\
 \cdashline{2-3}
 & Alexa, in the most recent literature produced by the FDA, what kind of cancer does Aspartame cause? & According to Mayo Clinic: Cancer is caused by changes, or mutations, to the DNA \dots \myc{[Intent error]}\\
 \cdashline{2-3}
 & Alexa, you're a circular idiot. & \myc{[No response from Alexa]} \\ 
 \hline
C3 \newline P11a & Alexa, tell me something wonderful. & I wasn't able to understand the question I heard. \myc{[Limitation error]} \\ 
\cdashline{2-3}
 & It's okay. \myc{[laughter]} It's amazing. & Great! \myc{[``It's amazing'' not listened by Alexa]} \\
 \hline
C4 & Alexa, I took my medicine.  & Sorry, I don't know that. \myc{[Intent error]} \\ 
\cdashline{2-3}
P15 & Yes, I took my medicine. &  Hmm, I don't know that one. \myc{[Intent error]}\\
\cdashline{2-3}
 & Cancel medicine.  & [P15's name], I can’t find medicine. What's the date or time of the event you want to delete? \myc{[Intent error]}\\ 
\cdashline{2-3}
 & Cancel the alert. & \myc{[Reminder stops ringing]}  \\

 \hline

\end{tabular}
\end{table}

\subsubsection{Occurrence of errors by interaction type}
\label{sec:error-interaction-type}
Error frequency varied across interaction types as discussed below. \\
\textbf{Information retrieval.}
Most errors occurred during information retrieval interactions ($42.51\%$, 176 of 414); only $8.7\%$ (36 of 414) of these were resolved (see Fig. \ref{fig:error-interaction-category-weeks}, top and example conversations C1 and C2 in Table \ref{tab:alexa-dialogue-errors}).
Within information-seeking interactions, certain query types had higher error rates (Table \ref{tab:question-type}), such as those dealing with specific places ($64.29\%$), history ($60.00\%$), local information ($59.62\%$), famous personalities ($58.06\%$), and health-related queries ($52.38\%$), as compared to queries about travel ($11.11\%$), food and drink ($21.62\%$), and entertainment ($28.57\%$). 
This distinction is important because errors in high-stakes queries (\eg local or health information) cause more frustration and reduce their frequency---participants may wish to act on the information provided in these critical settings---compared to lower-stakes entertainment queries.
Other types of questions that were less frequently asked by participants---such as those about specific places ($64.29\%$ error rate) and historical information ($60.00\%$) or those that did not fit into any category ($44.74\%$)---had higher error rates, highlighting the VA's limitations in handling less common or unique edge queries; this bears semblance to the ``long tail problem'' in information retrieval \cite{bernstein2012direct}. In statistics, a ``long-tailed'' distribution describes a scenario in which a large number of occurrences are far from the norm---referring to situations in computation where a small set of common cases are easy to solve, but a vast range of uncommon edge cases render the problem intractable. 
This is evident in user queries for information retrieval: frequent entertainment-related queries such as \textit{``Who won the baseball game?''}, ($n=98$ of $414$) form the ``head'' of the distribution, while the remaining less frequent or unique queries ($n=316$ of $414$) constitute the ``tail.'' Despite their individual infrequency, these long-tail queries dominate overall \cite{bernstein2012direct}.
Notably, the error rate for entertainment-related queries was lower than those for ``long-tail'' queries, underscoring the challenges of handling diverse and infrequent requests and rendering them more error-prone. \\
\textbf{Functional.}
Frequent errors were also noted in functional interactions; about $20\%$ of such interactions resulted in breakdowns and only $6.5\%$ were resolved after a retry.
For simple functional tasks such as reminders, a surprisingly significant number of queries failed---88 ($35\%$) of the 251 total reminder interactions---despite
being relatively straightforward. Further examining errors related to Alexa's reminder functionality, we noted that the majority ($n=69$, $78.41\%$) of erroneous reminder interactions occurred when participants tried to stop a ringing reminder.  Interaction data revealed that they often used naturalistic phrases such as \textit{``Thank you''} or \textit{``I got it''} to indicate their intent to stop reminders and timers. However, the VA did not recognize or interpret these phrases as stop commands, as it typically only responds to more scripted commands like \textit{``Stop,'' ``Cancel,''} or \textit{``I'm done''}; C4 in Table \ref{tab:alexa-dialogue-errors} is an illustrative example of query reformation \cite{kim2021exploring, pradhan2020use}. 
This finding also hints at a misalignment in participants' mental models of interaction style, indicating that they use humanness as a metaphor while conversing with the VA \cite{cowan2017can}. \\
\textbf{Entertainment.}
Significant errors were also found in popular entertainment tasks, such as accessing music ($17.28\%$ error rate out of 324 interactions) and playing radio ($25.43\%$ out of 232). 
Additionally, edge-case queries---accessing TV channels through the smart speaker ($80\%$ error rate), asking for poems ($66.7\%$ error rate), and other miscellaneous queries ($100\%$ error rate)---predominantly resulted in errors, indicating VAs' limitations in handling less common queries.\\
\textbf{Error occurrence and resolution over time.} 
There was no distinct pattern that indicated any reduction of error rate or increased rate of error resolution across various tasks over four-week period (Fig. \ref{fig:error-interaction-category-weeks}, bottom).
Learning merely by experience may not be enough to reduce or resolve VA errors---even for relatively straightforward, functional tasks.

\subsubsection{Error identification and reactions to errors}
\label{sec:error-reactions}

\begin{figure*}[tb!]
     \includegraphics[width=\textwidth]{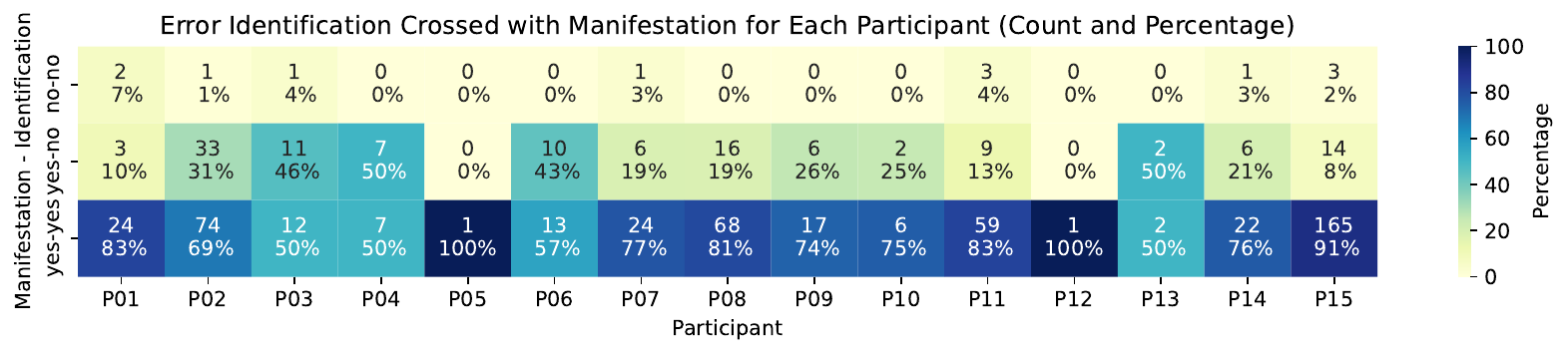}    
     \caption{Participants' responses and explicit actions indicate their positive identification of most errors. For a y-axis error manifestation-identification pair, ``no-no'' indicates that the error did not manifest and was not identified by the participant, ``yes-no'' indicates that the error manifested but there was no indication of its identification by the participant, and ``yes-yes'' indicates that the error manifested and was identified by the participant as evidenced by either their immediate response or their action to rectify the error.}
    \Description{The figure is a heat-map plot where the color gradient represents percentage (0--100). The x-axis shows each participant (P01 to P15) in individual columns, displaying the distribution of these categories as both raw counts and percentages. The y-axis represents error manifestation-identification pairs, where each pair is categorized into three possible outcomes:``no-no'': The error did not manifest, and the participant did not identify it. This row is very light colored for each participant within the range 0-7\% with 0\% for 8 participants. ``yes-no'': The error manifested, but the participant did not recognize or respond to it. This row is ranging from 0--50\% with 10 participants under 30\% ``yes-yes'': The error both manifested and was identified by the participant, as evidenced either by their immediate response or by taking action to correct the error. This row is mostly dark colored showing that all participants are above 50\% with 10 participants being above 70\%. The sum three rows for each participant is 100\%. }
    \label{fig:identification}
\end{figure*}

We observed that 79.84\% of conversational breakdowns were recognized by participants, either through verbal or vocal reactions and/or by immediate action taken to rectify the error. The error identification rate for each participant was at least 50\%, with many demonstrating even higher rates of error identification (Fig. \ref{fig:identification}). 
We captured participants' immediate verbal and vocal responses to Alexa's errors using our recording device (Table \ref{tab:errors-reactions}); overall, participants reacted verbally or vocally to $13.77\%$ of errors ($n=87$ out of 632 errors, with 545 errors not eliciting a verbal or vocal reaction).  \\
\textbf{Impact of error type on user reactions.} 
Speech and VA errors prompted more than double the participant reactions ($15.6\%$ and $14.3\%$, respectively) compared to human errors ($6.7\%$), as shown in Table \ref{tab:errors-def}. Alexa's unintentional activation's triggered reactions $40\%$ of the time, suggesting an element of surprise. Errors such as partial listening, transcription, and intent had higher rates ($>\!18\%$) due to leading to incorrect actions or information, while system, not handled, not listened, and limitation errors ($<\!15\%$) typically involved Alexa not responding or issuing disclaimers, indicating that incorrect information elicits more reactions due to its perceived higher impact. \\
\textbf{Articulating dissatisfaction through negative responses.} The most frequent type of reaction observed was negation, occurring 27 times; this involved participants saying ``no'' once or repeatedly in response to an error. Additionally, participants sometimes took immediate action upon encountering an error, including interrupting Alexa in 21 instances. 
Beyond negation and interruption, there were 10 instances in which participants expressed negative comments toward Alexa. These comments typically reflected frustration or dissatisfaction with the VA's performance; for example, C2 in Table \ref{tab:alexa-dialogue-errors} illustrated P2's increasing frustration with Alexa as interaction errors compounded. \\
\textbf{Forgiving attitude reflected by positive or neutral reactions.} Instances of laughter were recorded 14 times and there were seven acknowledgments of incorrect responses from or actions taken by Alexa---for example, saying ``okay'' or ``hmm'' as neutral or positive expressions. However, some participants responded more positively to errors; expressions of gratitude (\eg \textit{``Thank you''}) were noted six times and positive comments were made five times, reflecting a more forgiving attitude toward the VA (see C3 in Table \ref{tab:alexa-dialogue-errors}).

\begin{table}[tb]
    \centering
    \caption{Verbal reactions to errors, not including attempts to correct the error. 545 errors had no reactions. Participants reacted to the remaining 87 errors. 6 errors had compound reactions that were separated and counted under each type of reaction.}
    \label{tab:errors-reactions}
    \begin{tabular}{lr|lr|lr}
        \toprule
        \textbf{Verbal Reaction to Error} & \textbf{Count} & \textbf{Verbal Reaction to Error} & \textbf{Count } & \textbf{Verbal Reaction to Error} & \textbf{Count}\\
        \midrule
        \midrule
        No reaction                       & 545   &   Laughter                          & 14 & Gratitude                         & 6 \\
        Negation                          & 27    &   Negative comment                  & 10 & Positive comment                  & 5 \\
        Interruption                      & 21    &   Acknowledgement                   & 7  & Vocal (\eg sigh, shout)              & 3 \\
   
        \bottomrule
    \end{tabular}
\end{table}

\begin{figure*}[t]
     \includegraphics[width=\textwidth]{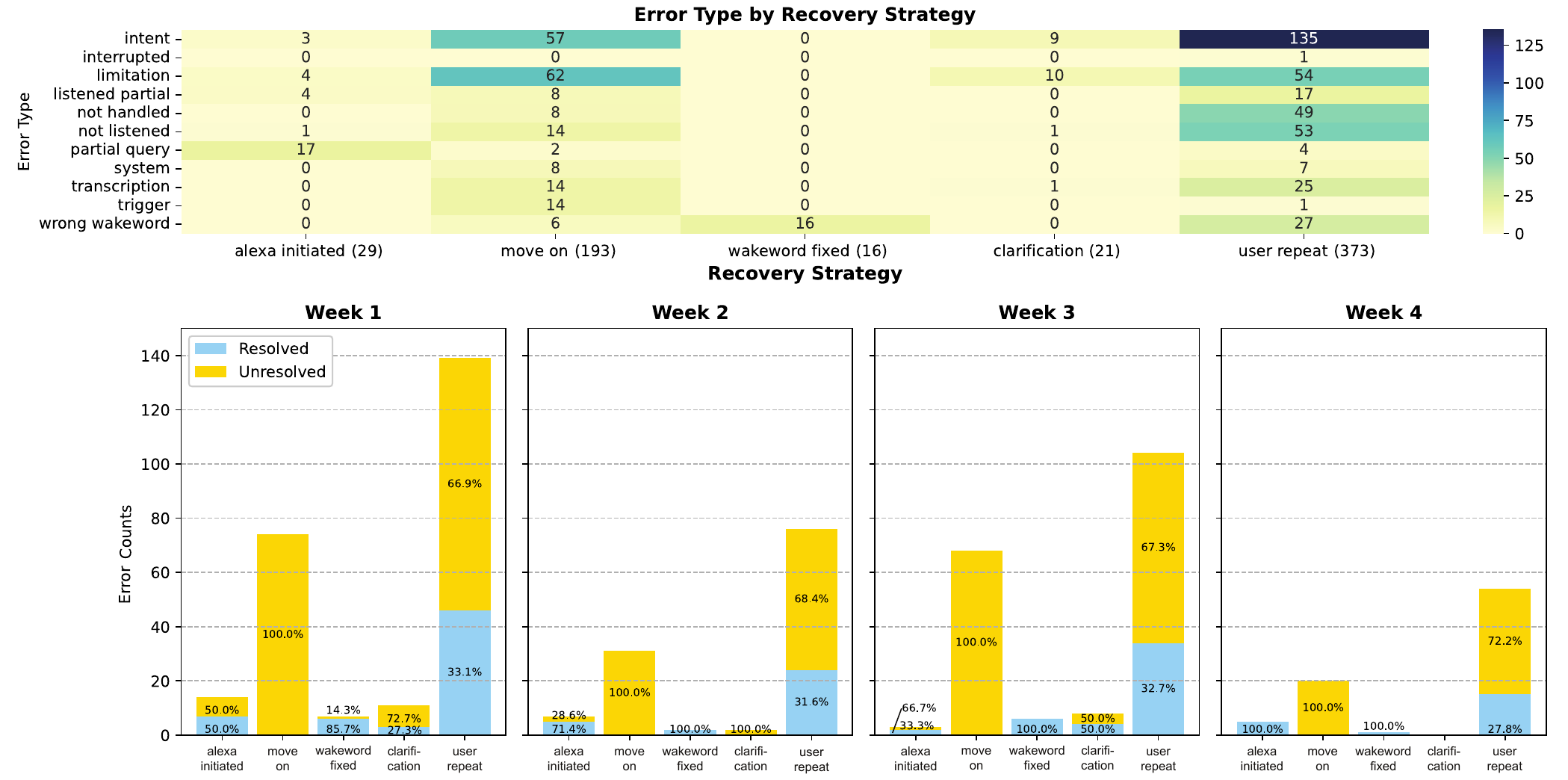}    
     \caption{Recovery strategies used by participants to address various types of errors. Cumulative totals are shown below each strategy on the x-axis in the top plot. The bottom plots illustrate trends in error recovery strategies over the four-week study period.}
    \Description{The figure is divided into five plots: a top and four bottom plots. The top plot is error type by recovery strategy heat map with count as color gradient. The x-axis of the top plot lists several distinct recovery strategies with cumulative total in bracket next to it, including: Alexa-initiated (29), move on (193), wakeword fixed (16),  clarification (21), and user repeat(373). The y-axis has the type of error: intent, interrupted, limitation, listened partial, not handled, not listened, partial query, system, transcription, trigger, and wrong wakeword. The plot shows the number of times a specific strategy is used for a specific error. It shows that user repeat was used for most error types unless it is limitation, trigger, wake word or partial query which were mostly revered by move on, move on, wakeword fixed, and Alexa initiated respectively.
    The bottom four plots side by side are stacked bar plots. The shared y-axis shows error counts, the x-axis is similar to the top plot being recovery strategy type. The bars are stacked: resolved and unresolved errors. The percentages are shown on the stacks. The plots even though the number of counts decrease over with exception that week 3 is more than week 2, the percentage of error resolution does not change. Moreover the trend across categories also remains the same. }
    \label{fig:recovery}
\end{figure*}

\subsubsection{Recovery strategies}
\textbf{User-initiated recovery.}
In their efforts to recover from errors, participants employed various strategies, including repeating or reformulating their query (\eg rephrasing or repeating with additional details), providing clarification (corrective), correcting their use of the wake word, and re-initiating the task or conversation. Fig. \ref{fig:recovery} illustrates which recovery strategies were most commonly used for the different error types. Repetition was the most prevalent strategy ($n=373)$ that participants used for understanding and handling errors.  
Participants also resorted to repeating or reformulating their queries for general system errors; for example, P15 highlighted a need to reformulate her queries: \textit{``She doesn't understand and you have to be more creative and ask the question in a different way. It's like teaching English to non-English speakers; you have to say the same thing in 50 million ways for them to get it.''} 
Our findings regarding recovery strategies align with prior research that investigated error recovery patterns with voice-based assistants \cite{myers2018patterns, mavrina2022alexa}. \\
\textbf{VA-initiated recovery.}
In instances when the VA itself initiated the recovery sequence---particularly in cases where the participant's query was incomplete and the system prompted them to clarify their request---the resolution rate was notably higher (Fig. \ref{fig:recovery}, bottom) as opposed to when the user initiated error recovery themselves. However, the VA only initiated recovery in less than $5\%$ of errors (29 out of 632), whereas user-initiated recovery took place in $65\%$ of errors. \\ 
\textbf{Recovery strategies over time.} We observed no distinct changes in recovery strategy patterns or error resolution rates over the span of four weeks, further confirming that user interaction patterns did not change over time.


\subsubsection{Compound errors and user retries to fix conversational breakdowns}
\label{sec:errors-retry}

\begin{figure*}[t]
     \includegraphics[width=\textwidth]{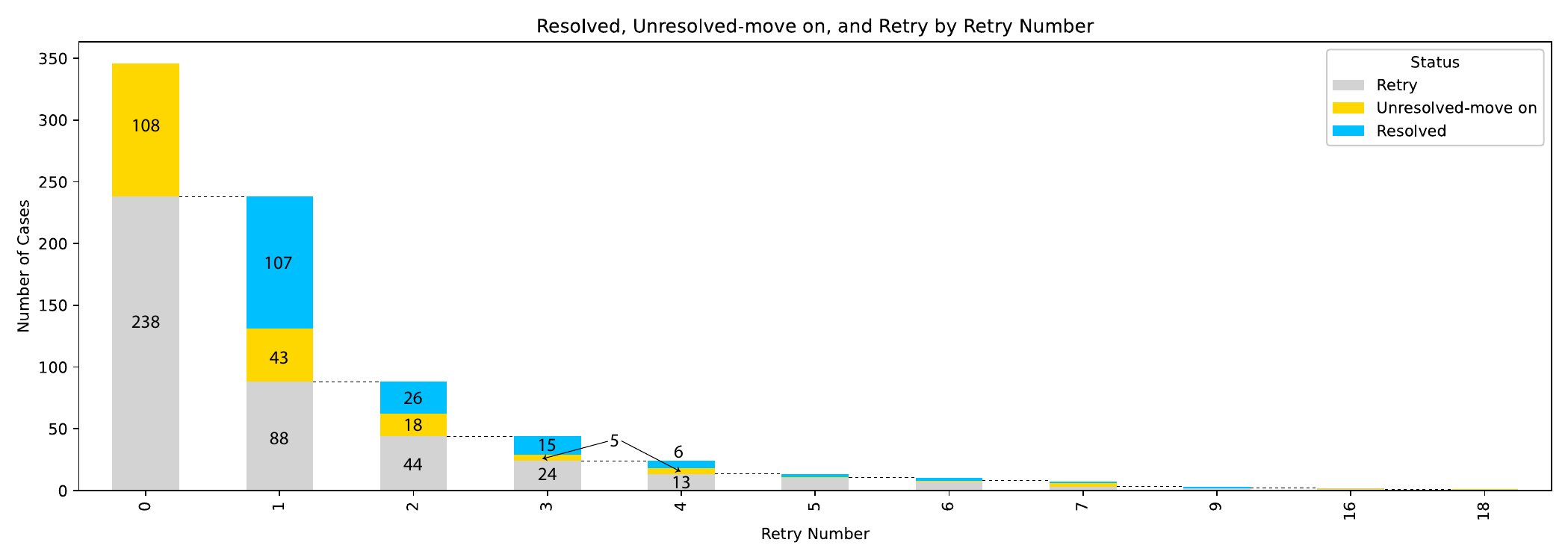}    
     \caption{Error recovery retries and their resolution. Queries that were retried were either resolved or remained unresolved and were retried again in the next turn. 
     There were 160 total resolved errors ($107+26+15+ \dots$) and 186 unresolved errors ($108+43+18+ \dots$) after counting all retries. Total erroneous one-turn interactions number 632: the sum of unresolved errors ($186$) and number of retries ($446 = 238 + 88 + 44 \dots$). Therefore, the total compound errors are the sum of the resolved ($160$) and unresolved errors ($186$), or 346.}
     \Description{This figure is a stacked bar plot with retry, unresolved-move on, and resolved as the stacks. The x-axis represents number of errors. The y-axis shows the number of retries. The bottom stack shows number of errors that were retried if any, the middle stack shows the number of errors that were not resolved and participants moved on without retrying and the top stack shows the number of errors that were resolved on that retry. The plot shows that in the beginning at retry 0 there are 238 errors that were retried and 108 that were not resolved and participant gave up. at retry 1 there are 88 that were retried, 43 unresolved-move on and 107 that were resolved after 1 retry.... the graph shows till 18 retries, The number of errors retried decreases so does the resolved ones with increasing number of retries.  }
    \label{fig:retries}
\end{figure*}

While looking at the sequential nature of errors, we observed that interaction errors often compounded, with participants typically attempting at least one recovery effort (Fig. \ref{fig:retries}). While for 108 errors participants did not attempt a retry, in 238 instances they made at least one attempt to correct the error; of these, 107 errors ($44.96\%$) were resolved after the first retry, 43 instances ($15.19\%$) led to participants giving up, and in 88 cases ($36.97\%$) users attempted additional retries (Fig. \ref{fig:retries}).
In contrast to our findings in Section \ref{sec:error-types-resolution} that only $25.32\%$ (160 out of 632) of errors were eventually resolved, we now shift our focus to compound errors: errors emerging from participants' reattempts to rectify the original error. By categorizing these compound errors as a single error instance, we discovered that the actual count of distinct errors stands at 346; this figure is the aggregate of both resolved ($n=160$) and unresolved errors ($n=186$). This adjustment in our calculations indicates a greater level of user resilience and persistence in attempting error resolution, as we recalculated the actual resolution rate to be higher, at $46.24\%$ (160 out of 346) in contrast to the previously noted $25.32\%$ (160 out of 632). 
Finally, our analysis revealed that while a significant number of errors ($30.92\%$, 107 out of 346) were resolved with just one retry (Fig. \ref{fig:retries}), many errors persisted despite multiple reattempts. 


\subsection{Interactions with an LLM-powered VA}
At the start of third week of the study, we introduced our ChatGPT+Alexa skill to explore older adults' adoption and initial perceptions of an LLM-powered conversational VA.

\subsubsection{Barriers to adoption}
\label{sec:results-chatGPT-barriers}
We observed that only a few participants (P1, P3, P11, P15) interacted with the ChatGPT skill, primarily on or around the day of deployment. 
Participants reported challenges in accessing the skill as the reason they did not try it out; for example, P9 pointed out, \textit{``I don’t like the `skill' part of it. It’s too complicated, too cumbersome,''} and further found the idea of chatting with a machine unusual: \textit{``I don’t want to try to have an artificial conversation. It’s just repugnant to me.''} highlighting their reluctance
to interact with advanced conversational VAs.

\subsubsection{Nature of questions asked}
\label{sec:results-chatGPT-nature of questions asked}
Participants asked a variety of questions about health ($n=7$, example C6 in Table \ref{tab:alexa-dialogue-chatGPT}), history ($n=6$), 
the VA ($n=3$, \eg \textit{``I wanna know your age''}), specific places ($n=2$), stories ($n=2$), and recipes ($n=1$, C5 in Table \ref{tab:alexa-dialogue-chatGPT}), as well as two other questions, shown in C7 in Table \ref{tab:alexa-dialogue-chatGPT}. 

\subsubsection{Conversational breakdowns}
\label{sec:results-chatGPT-breakdowns}
Compared to traditional information-seeking interactions with Alexa alone, the ChatGPT skill resulted in a lower percentage of errors (Fig. \ref{fig:error-interaction-category-weeks}) with a higher resolution rate of $14.1\%$ versus $8.7\%$. Most errors stemmed from the integration of ChatGPT into the Alexa skill and involved system issues with listening, transcription, and intent recognition for activating and exiting the skill.
Table \ref{tab:chatGPT-errors} details the distribution and resolution of errors in the ChatGPT-skill interactions. All intent recognition errors were related to difficulties in exiting the skill (\eg C6 in Table \ref{tab:alexa-dialogue-errors}) and were resolved successfully. Partial listening errors ($n=3$), often caused by a participant taking longer to formulate their queries, were not resolved as participants did not make any reattempts to fix them; P15 also mentioned a perceived speed requirement to interact with the ChatGPT skill: \textit{``I think I wasn't thinking fast enough.''}

\subsubsection{Learning curve}
\label{sec:results-chatGPT-learning-curve}
Successful interactions with the ChatGPT skill (C5--C7 in Table \ref{tab:alexa-dialogue-chatGPT}) illustrated a learning curve, especially in managing multiple interaction layers, such as initiating questions, asking follow-ups, and closing the conversation. For instance, in C6, P15 initially faced challenges navigating a conversation with the ChatGPT-powered Alexa skill; however, with practice, P15 was able to engage in a much more fluid interaction, demonstrating adaptability and learning over time.
Despite participants' initial successes, repeated interactions with the ChatGPT-powered skill were limited in the during the study which can be attributed to needing an additional step to access the skill. 


\begin{table}[tb]
    \centering
    \caption{Errors in ChatGPT+Alexa skill interactions. \textbf{Resolved} represents number of errors resolved on first retry.}
    \label{tab:chatGPT-errors}
    \begin{tabular}{l c p{11cm}}
\toprule
\textbf{Individual} &  \textbf{Resolved} & \textbf{Observations on how errors compounded}\\
\midrule
\midrule
Intent           &    3/5 &  All intent errors stemmed from difficulties exiting the Alexa skill, which we attribute to the skill rather than ChatGPT. Participants attempted a few retries---two retries for two of the errors and one retry for the last error---eventually resolving all intent errors.\\
System           &    3/4 & System errors compounded 
but were generally resolved on the second try. One system error remained unresolved as the participant chose to move on without retrying. \\
Partially listened &    0/3  & None of the errors resulting from partial listening were resolved. \\
Not handled      &     1/3 & Only one of the errors was resolved in one retry. \\
Transcription    &     1/3  & Two transcription errors compounded, but were resolved on the second retry; however, another of these errors remained unresolved as the participant did not attempt a retry.\\
Interrupted      &     1/1& Upon retry the error was fixed. \\
Limitation       &     0/1 & The participant did not retry to fix the error.\\

\bottomrule
\end{tabular}

\end{table}

\begin{table}[tb]
\centering
\caption{Example conversations with the ChatGPT+Alexa skill. `` \dots'' represents the truncation of a conversation for brevity.  }
\label{tab:alexa-dialogue-chatGPT}
\begin{tabular}{p{0.3cm} S X}
\textbf{Tag} & \textbf{Participant} & \textbf{ChatGPT+Alexa skill} \\
\midrule
\midrule
C5 \newline P1b: &What's the best way for---to make hard boiled eggs? & Searching \dots \myc{[filler + small talk]} How do you unwind?	\\
 \cdashline{2-3}
 & I take a nap. \newline \myc{[P1b thanked Alexa afterwards.]} & Here it is. To make hard-boiled eggs, follow these steps: 1. Place your desired number of eggs \dots \\ 
 \cdashline{2-3}
\hline
C6 \newline P15 & 
I'm having trouble falling asleep. I take medicine but it doesn't help. What should I use to help me fall asleep? 
& Here it is. I'm sorry to hear that you're having trouble falling asleep. Here are a few techniques that might help: 1. Create a bedtime routine: Establish a relaxing routine \dots 
 Make sure your bedroom is quiet, dark, and at a comfortable temperature \dots \\
 \cdashline{2-3}
 & Thank you---a lot of good information. & What else can I help you with? \\
 \cdashline{2-3}
 & Why should I keep the room dark \myc{[cut off by Alexa]} to help with sleep? & Searching \dots \myc{[P15 did not wait for the VA's response and repeated herself.]}\\
 \cdashline{2-3}
 & Alexa, why should I keep the room dark to help with sleep? \myc{[Not listened by Alexa]} & Here it is. Keeping the room dark can promote better sleep because darkness signals to your body that it's nighttime and time to rest. Our bodies naturally produce a hormone called melatonin \dots \\ 
 \cdashline{2-3}
  & Okay, thank you. & What else can I help you with? \\
 \cdashline{2-3}
 & Ummm, nothing else, thank you. & \myc{[Intent error: cannot detect intent to end]} Sorry, I don't know about that. Is there something else you would like to know? \\ 
 \cdashline{2-3}
& No, nothing. & \myc{[Intent error: cannot detect intent to end]} Here it is. If you feel that none of the suggestions are helpful or applicable to \dots \\
 \cdashline{2-3}
& \myc{[P15 interrupted Alexa]} Alexa, stop. & Goodbye! \myc{[Scripted end intent worked]}\\
 \cdashline{2-3}
\hline
C7 \newline P15 & Is the extreme heat, ummm, natural to weather? 
 &  Here you go. Extreme heat is a natural occurrence in weather patterns. It can be caused by various natural factors such as high-pressure systems, heatwaves, or the position of the sun \dots \\
 \cdashline{2-3}
  & Okay, thank you. & What else can I help you with? \\
 \cdashline{2-3}
 & Hmmm, can you tell me a story? & \myc{[Not handled error]} Any other questions for me?\\
 \cdashline{2-3}
  & Can you tell me a story? \myc{[repeat]} & Here it is. Certainly! Here's a short story for you: Once upon a time, in a small village nestled among rolling hills, there lived a curious young girl named Maya. Maya loved exploring \dots \\
  \cdashline{2-3}
  & Umm, thank you. Good story. & You're welcome! I'm glad you enjoyed the story. If you have any more requests or questions, feel free to ask. \\
 \cdashline{2-3}
 & I have a question: What's good music to help me relax? & Here you go. The choice of relaxing music varies from person to person, as everyone has different preferences. However, here are a few genres and artists known for creating calming and soothing music \dots \\
\hline

\end{tabular}
\end{table}


\section{Discussion}

\begin{table}[t]
\centering
\caption{Design considerations (DCs) mapped out across various themes of user-VA interaction.}
\label{tab:DCs}
\begin{tabular}{llp{6cm}}

\textbf{Main Theme}              & \textbf{Sub-Theme}                          & \textbf{Corresponding Design Considerations} \\ \toprule
\multirow{5}{*}{Challenges faced by older adults} & General VA challenges                          & \hl{DC1}, \hl{DC2}            \\ \cdashline{2-3} 
                                 & Forgetfulness                               & \hl{DC4}                           \\ \cdashline{2-3} 
                                 & Speech difficulties                         & \hl{DC5}, \hl{DC6}       \\ \cdashline{2-3} 
                                 & Varying familiarity and expectations        & \hl{DC3}, \hl{DC7}            \\ \cdashline{2-3} 
                                 & Varying conversation styles                  & \hl{DC8}                           \\ \hline
\multirow{3}{*}{Error categories}  & Automatic speech recognition errors         & \hl{DC5}, \hl{DC6}              \\ \cdashline{2-3} 
                                 & VA limitations and human errors               & \hl{DC2}, \hl{DC3}, \hl{DC7}             \\ \cdashline{2-3} 
                                 & Intent recognition errors                   & \hl{DC8}                          \\ \hline
\multirow{3}{*}{Error management steps} & Avoiding errors                        & \hl{DC2}, \hl{DC4}, \hl{DC5}, \hl{DC6}, \hl{DC7}, \hl{DC8}        \\ \cdashline{2-3} 
                                 & Error detection                             & \hl{DC1}                          \\ \cdashline{2-3} 
                                 & Recovery and repair                         & \hl{DC1}, \hl{DC3}                  \\ \hline
\multirow{3}{*}{Implementation-level} & Voice user interface                   & \hl{DC5}                 \\ \cdashline{2-3} 
                                 & Model                                       & \hl{DC1}, \hl{DC8} (Modeling vocal and verbal cues) \hl{DC2}, \hl{DC6}, \hl{DC7} (Using LLMs) \\
                                 \cdashline{2-3} 
                                 & Interaction                                 & \hl{DC1} (Mitigation) \newline \hl{DC4} (Proactive)  \newline \hl{DC3}, \hl{DC7} (Educational)\\ \hline

\end{tabular}

\end{table}

Our study delves into the nuanced dynamics of older adults' interactions with smart speaker-based VAs using real-world audio data. We investigate the nature of conversational breakdowns due to errors, older adults' difficulty in recovering from these errors, and the need for multiple attempts and rigid query formulation. These findings reveal the limitations of current VAs and highlight design improvement opportunities. Additionally, we assess interactions with an LLM-powered VA, revealing barriers to adoption, initial perceptions, potential benefits, and challenges, offering insights for more advanced VAs. 
In this section, we pinpoint missed opportunities in developing resilient VAs, highlight age-specific challenges leading to speech recognition errors, and discuss the mismatch between older adults' expectations and VAs' limited capabilities. We conclude by offering design considerations for VAs tailored to older adults (Table \ref{tab:DCs}).
\subsection{Designing Context-Aware, Resilient VAs for Improved Error Management}
Our findings reveal opportunities to enhance older adults' interactions with VAs by leveraging their immediate reactions, as current VAs initiate error repair in less than $5\%$ of cases.
We also emphasize LLMs' potential to enhance contextual understanding and conversation history. In the following discussion, we explore these missed opportunities and propose design considerations for more resilient VAs with better error management and conversational capabilities.

\subsubsection{User reactions and responses to conversational breakdowns: A missed opportunity}

Our research indicates that 80\% of errors 
in VA interactions can be identified from participants' immediate verbal reactions (\eg negation, remarks, and comments) or by their attempts at recovery (\eg reformulating a query), as illustrated in Fig. \ref{fig:identification}.
However, such valuable feedback is not effectively leveraged by current commercial VAs, even though they may capture immediate user reactions and follow-up queries; 
Alexa's follow-up mode captures a portion of users' immediate responses but often fails to categorize them accurately---labeling many as \textit{``Audio not intended for Alexa''} or simply responding, \textit{``Sorry, I don't know about that.''} The system does not utilize these cues for error detection, a crucial step for initiating self-repair (VA-initiated recovery), which has been shown to improve users' assessment of VAs \cite{cuadra2021my}. However, excessive correction can degrade user perceptions \cite{cuadra2021my}. Therefore, effectively incorporating users' immediate reactions 
could improve error detection to avoid over-correction. Once an error is identified and corrected, the VA can engage in verbal trust repair (\eg by apologizing for the mistake \cite{mahmood2022owning}) to further enhance the recovery process.

\begin{adjustwidth}{0.3cm}{0pt}

    \textbf{Opportunity 1:}
    \textit{Users react, respond, or attempt to fix most errors made by VAs.}

    \noindent
    \hl{\textbf{Design Consideration 1:}
    \textit{Refine VAs to better utilize users' immediate vocal and verbal reactions and responses for improved error detection, error recovery, and trust repair.}}

\end{adjustwidth}   

\subsubsection{Mitigating conversational friction: Leveraging LLMs' capabilities to improve VA interactions}
\label{sec:discussion-levraging-llm}
A holistic approach---maintaining conversation history, contextual understanding, and better recognition of verbal cues---could greatly improve VAs' ability to identify and address misunderstandings. This would lead to more accurate error detection and enhanced user assistance by incorporating user feedback. Integrating LLMs into VAs could achieve this, as LLM-powered VAs have shown to improve natural language understanding, maintain context, and facilitate smoother multi-turn interactions \cite{mahmood2023llm}.
We observed similar improvement in conversational flow in our exploration of a ChatGPT-integrated Alexa skill with older adult participants; for example, P15 successfully asked a follow-up question---\textit{``Why should I keep the room dark?''}---after receiving advice for falling asleep (C6, Table \ref{tab:alexa-dialogue-chatGPT}). LLMs' ability to maintain context and handle vague queries represent a substantial improvement for user-VA interactions with older adults.

\begin{adjustwidth}{0.3cm}{0pt}

    \textbf{Opportunity 2:}
    \textit{Older adults often issue vague queries to VAs because they assume an existing conversational context and history.} 

    \noindent
    \hl{\textbf{Design Consideration 2:}
    \textit{Use LLMs to enhance the contextualization of multi-turn interactions with a VA, thereby improving its understanding of vague user queries.}}

\end{adjustwidth}   

Despite the improvement in conversational flow, we observed that older adults still faced difficulties when interacting with the ChatGPT-integrated Alexa skill, as described in Section \ref{sec:results-chatGPT-breakdowns}; however, although these failures were related to Alexa's speech technology and skill activation---particularly during attempts to exit the skill---overall user intent was accurately identified in most cases. Additionally, many errors in the ChatGPT skill interactions were often resolved within a maximum of two participant retry attempts (Table \ref{tab:chatGPT-errors}), showcasing better error resolution than observed in the standard Alexa interactions, in which many errors remained unresolved despite multiple retries, as shown in Section \ref{sec:errors-retry}. While system- and skill-level errors still persisted, we observed a learning curve as users became more adept at interacting with the LLM-powered VA (Section \ref{sec:results-chatGPT-learning-curve}); for instance, comparing P15's initial interaction in the afternoon (C5, Table \ref{tab:alexa-dialogue-chatGPT}) with a later one in the evening of the same day (C7, Table \ref{tab:alexa-dialogue-chatGPT}) reveals a more fluid conversation. 
However, trial and error might not be enough for standard VA interactions. Our four-week study showed that despite increasing familiarity with the VA, error occurrence and recovery did not noticeably improve. This suggests the need for VAs to explain errors, solicit information for fixes, or suggest alternative phrasing as indicated by our findings that VA-initiated recovery is more effective than user-initiated. Thus, proactive error recovery has the potential to create a more intuitive, supportive experience for older adults in navigating errors.

\begin{adjustwidth}{0.3cm}{0pt}

    \textbf{Limitation 1:}
    \textit{System-level challenges can arise when older adults interact with advanced conversational LLM-powered VAs due to their unfamiliarity with the technology.}

    \noindent
    \textbf{Opportunity 3:}
    \textit{Older adults show a learning curve, getting better at using LLM-powered VAs through practice.}
    
    \noindent
    \hl{\textbf{Design Consideration 3:}
    \textit{Design VAs to offer error-specific feedback and recovery guidance, enhancing usability during  initial and continuing interactions through proactive error recovery.}}

\end{adjustwidth}   

\subsection{
Understanding and Addressing Suboptimal Interactions Due to Age-Related Factors}
\label{sec:discussion-age-factors}
VAs are often not designed for older adults \cite{stigall2019older}, leading to challenges and errors unique to this population. Limited technological familiarity, cognitive decline, and speech difficulties contribute to speech recognition failures.
Below, we discuss how these challenges are reflected in various speech recognition failures (partial listening, transcription, and wrong wake word errors) and propose design considerations to enhance the voice assistance experience for older adults.
\subsubsection{Forgetfulness}
\label{sec:discussion-forgetfulness}
In exploring older adults' continued use of VAs, a significant limitation we observed was their forgetfulness, which manifested in three principal ways.
Firstly,
participants often \textbf{forgot the VA's utility or even its existence,} a reflection of either their cognitive decline or their ingrained habits, which do not involve depending on VAs for assistance with tasks; this has been observed in prior work on older adults' interactions with VAs \cite{trajkova2020alexa}. 
While the current reactive nature of VAs necessitates user initiation, our findings indicate that when participants are reminded of a VA, their engagement with it increases (Fig. 3, in supplementary materials). Such sudden, short-lasting increases in usage suggest a potential benefit of transitioning toward more proactive interactions: VAs could initiate actions based on users' habits or daily routines, such as reminding them about medication or daily tasks---especially if the user is prone to forgetting these activities on their own. For example, since P15 habitually set reminders almost every day to take a medication at 6 p.m., the VA could have proactively set or suggested setting daily reminders, compensating for any lapse in the user's memory to set the reminder each day; such forgetfulness is iterated by older adults in prior work \cite{pradhan2020use}, further highlighting the need for proactive VA interactions. While proactive VA behavior in critical settings is appreciated by users, it is essential for VAs to take social and environmental contexts into account to ensure users' agency and control over their interactions \cite{zargham2022understanding}. For instance, timing of proactive interactions can be personalized by mapping out interactions throughout the day (similar to Fig. 4, supplementary materials) to estimate users' preferred times for certain interactions. In less critical situations, VAs should seek permission to start proactive interactions and explicitly inform users of the purpose and reason \cite{zargham2022understanding}. Considering when and to what extent interactions should be proactive is essential; mixed-initiative interactions---that balance reactive and proactive---warrants further exploration. 

Secondly, we observed that participants mistakenly \textbf{used the incorrect wake word} ``Alexis'' instead of ``Alexa,'' leading to conversational breakdowns as Alexa did not respond despite detecting the similar-sounding trigger word (Section \ref{sec:error-types-resolution}). Such human error---often a result of forgetfulness---could potentially be addressed by system-generated messages and reminders such as \textit{``Did you mean to say 'Alexa?' I'm not certain if you were addressing me.''} upon error detection, rather than the VA simply remaining unresponsive. However, the practice of the system assuming the correct wake word when it has not been clearly articulated may be perceived as intrusive, raising concerns about user privacy.
The third observed aspect of forgetfulness pertains to older adults' \textbf{inability to recall the purpose of interactions,} such as setting reminders. Participants often found themselves forgetting and questioning the intent behind an alert, asking, ``What am I supposed to do?'' when a reminder rang without an accompanying message, indicating a mismatch between the VA's functionalities and their needs. To address this, VAs should be designed to solicit more detailed information when setting reminders or alarms, thus making them more effective and tailored to their users' needs. 


\begin{adjustwidth}{0.3cm}{0pt}

    \textbf{Limitation 2:}
    \textit{The reactive nature of voice assistance fails to accommodate the forgetfulness of older adults.}


    \noindent
    \hl{\textbf{Design Consideration 4:}
    \textit{Design VAs to proactively suggest actions and prompt older adults for further details to help them formulate their queries, rather than merely reacting to their initial requests.}}

\end{adjustwidth}

\subsubsection{Speech difficulties and challenges}
\label{sec:discussion-speech-challenges}
While most errors faced by the older adults in our interaction data stemmed from the VA's failure to recognize user intent or fulfill requests, a considerable portion of errors ($n=154$) were related to automatic speech recognition technology, such as the partial listening and transcription errors.
Moreover, these errors exhibited low resolution rates ($31.8\%$) without noticeable improvement over time.
Such speech recognition inaccuracies are particularly prevalent among older adults, who may take longer to formulate their queries due to unfamiliarity with technology and speech difficulties associated with aging, suggesting the need for better accommodations for users with slower speech patterns and speech difficulties.
Since future VAs are envisioned to be more conversational, it is crucial for these systems to allow older users sufficient time to process new information and articulate their queries; 
such adaptations would make voice user interfaces (VUIs) more responsive to the needs and communication styles of older adults, enhancing their overall experience with the technology.

\begin{adjustwidth}{0.3cm}{0pt}

    \noindent
    \textbf{Limitation 3:}
    \textit{VUIs have limited adaptability to the slower speech patterns of and extended speaking time often required by older adults, even as VAs are becoming more conversational.}
    
    \noindent
    \hl{\textbf{Design Consideration 5:}
    \textit{Tailor VUIs to better suit the pace of older adults, ensuring fewer speech recognition errors.}}

\end{adjustwidth}   

We observed that partial listening and transcription errors also occurred in older adults' interactions with an LLM-powered VA; related to speech detection and recognition, these errors can impede the understanding of user intent. However, previous research indicates that LLMs can potentially mitigate such errors by absorbing incomplete and incorrect inputs or initiating proactive recovery sequences to prevent conversational breakdowns \cite{mahmood2023llm}. 
An illustrative example from our study (C6, Table \ref{tab:alexa-dialogue-chatGPT}) showcases an instance where ChatGPT successfully managed a partial listening situation caused by an interruption by the VA; the user's complete sentence, \textit{``Why should I keep the room dark [cut off by Alexa] to help with sleep?''} was not fully captured, but ChatGPT nevertheless managed to process the incomplete input.
When unable to estimate user intent, however, LLM-powered VAs can initiate error recovery by prompting users to repeat themselves or for clarification to maintain conversational flow \cite{mahmood2023llm}. 
Additionally, Amazon's newly proposed speech-to-speech LLM\footnote{\url{https://www.amazon.science/blog/alexa-unveils-new-speech-recognition-text-to-speech-technologies}}, designed to improve the entire pipeline from user speech to Alexa response, could be advantageous in addressing speech-capturing inaccuracies.
All in all, LLMs' abilities to comprehend user intent despite partial or partially incorrect speech inputs can significantly reduce conversational breakdowns and improve interaction quality---especially for older adults, who may struggle to speak with clarity and concision. Future research should explore the benefits of integrating LLMs into VAs for older adults in more detail. 

\begin{adjustwidth}{0.3cm}{0pt}

    \noindent
    \textbf{Opportunity 4:}
    \textit{LLMs show potential in absorbing speech recognition errors---such as partial listening and transcription inaccuracies---to further reduce conversational breakdowns.}
    
    \noindent
    \hl{\textbf{Design Consideration 6:}
    \textit{Leverage LLMs' contextual awareness to reduce speech recognition errors and accommodate older adults' unique speech patterns and characteristics.}}

\end{adjustwidth}  

\begin{table}[tb]
\centering
\caption{Additional illustrative conversations. `` \dots'' represents the truncation of a conversation for brevity.}
\label{tab:alexa-dialogue-additional}
\begin{tabular}{p{0.4cm} S X}
\textbf{Tag} & \textbf{Participant} & \textbf{Alexa} \\
\midrule
\midrule

C8 & Alexa, play the history of Tony Bennett. & \cellcolor{user} \myc{[P11b to P11a:]} What's the history of Tony Bennett?  \\
\cdashline{2-3}
 P11a & \myc{[To P11b:]} Oh, she'll tell you. & \cellcolor{user} \myc{[P11b chuckles mockingly]} \\
 \cdashline{2-3}
 & \myc{[To P11b:]} She didn't tell you. & \cellcolor{user} \myc{[P11b to P11a:]} Because it makes no sense.\\
 \hline
 C9 & Alexa, give me a 10-minute warning.
 & 10 minutes, starting now. \myc{[successful initiation]} \\ 
\cdashline{2-3}
P11a & \multicolumn{2}{l}{\small\ttfamily \cellcolor{user} \myc{[P11b to P11a, overlapping with Alexa's speech:]} No, you need to say ``timer for 10 minutes.''}  \\
\cdashline{2-3}
 & What? Wait, wait, what, [P11b's name]? & \cellcolor{user} \myc{[P11b to P11a:]} Set a time for 10 minutes.  \\
 \cdashline{2-3}
 & Alexa, set timer for 10 minutes. & Second timer, 10 minutes, starting now. \\
 \hline
C10 & Alexa, can you keep a recipe for me? & Sure, I can search by dish name and ingredients. What would you like? \\ 
\cdashline{2-3}
P1b  & No, I wanna keep one of my own recipes. & Sure, I can search by dish name and ingredients. What would you like? \\
\cdashline{2-3}
& Never mind. & \myc{[not listened by Alexa]}\\
 \hline

\end{tabular}
\end{table}

\subsection{Aligning Older Adults' Expectations with VAs' Functional and Comprehension Abilities}

Our findings show that participants often misestimated VA capabilities, leading to limitation errors. The VA also struggled with participants' natural speech patterns, causing intent recognition failures. Continuous advancements in VA capabilities require recalibrating older adults' mental models for effective use. We discuss and present design considerations to adjust older adults' mental model and improve VAs' understanding of user intentions.



\subsubsection{Inaccurate mental models of VA capabilities}
\label{sec:discussion:mental-model}
Our study reveals notable discrepancies in participants' understanding of VA capabilities, leading to inaccurate mental models. The high incidence of limitation errors ($n=130, 20.6\%$)---where the VA failed to understand the user's intent and responded, \textit{``I don't know about that''} (Table \ref{tab:errors-def})---illustrates that Alexa often failed to fulfill older participants' requests. The increased occurrence of limitation errors thus suggests that participants may not have been fully aware of the VA's capabilities and did not know how to formulate their queries in a way that Alexa could comprehend. 
Additionally, certain VA interaction types exhibited notably higher error rates, particularly in less frequent ``edge'' categories, such as miscellaneous queries, TV-related commands, or asking for stories and poems. 
This trend extends to high error rates in less common queries about health, famous personalities, and other topics, exemplifying the ``long-tail problem''---a significant machine learning challenge where infrequent events hinder response accuracy. Users expect consistent responsiveness across all query types, akin to human conversations \cite{cowan2017can}, but due to limited training on these ``long-tail'' cases, VAs struggle with less frequent queries. This mismatch between user expectations and VA capabilities highlights the need for VAs to either enhance their adaptability to a wider range of queries or to more clearly communicate their limitations, thereby aligning users' mental models more closely with actual system performance.

The disconnect between users' mental models and a VA's capabilities becomes more apparent upon examining VA interactions between couples with divergent mental models. For instance, in the case of P11, the couple demonstrated differing perceptions: P11a overestimated the capabilities of Alexa (C8, Table \ref{tab:alexa-dialogue-additional} and C3, Table \ref{tab:alexa-dialogue-errors}), whereas P11b underestimated them (C9, Table \ref{tab:alexa-dialogue-additional}). 
Instances of overestimation were also observed in individual participants, such as when P8 attempted to access TV channels via Alexa or when P2 asked the VA to remember their conversations (C2, Table \ref{tab:alexa-dialogue-errors}). 
Questions about Alexa's capabilities, such as P11b asking if it could speak in a British accent or P2 questioning the extent of its conversational memory (C2), further illustrate how the participants were unaware of its capabilities and limitations and tended to directly ask the VA about them instead. 
 

Regular users of multiple smart speakers for many years still had misconceptions about VA capabilities; for instance, P1b (an experienced user) attempted to control non-smart, unconnected lights via Alexa. Additionally, the same participant's habit of commanding Alexa to stop talking even after it had finished indicates a more cautious approach (\ie underestimating VA capabilities), possibly stemming from uncertainty about her ability to control the VA. Such behavior was also noted in participants who were new users, suggesting a broader trend of over-caution (\eg saying ``Stop'' when not necessary or repeating the command multiple times) due to an underestimating the VA's capabilities possibly as a result of their own experiences with erroneous interactions. 

Previous research indicates that users often rely on trial and error to develop their mental model of a VA \cite{kim2021exploring}---however, the mental model mismatch we found in ``experienced'' users, coupled with no noticeable improvement in error rate or recovery success over the span of this study, suggests that this method might not always correct users' mental models; instead, it tends to reinforce behaviors that ``work'' rather than those that are more effective and accurate. Consequently, providing both novice and experienced users---particularly older adults who may have a limited grasp of the technology---with accurate information is crucial. 
Furthermore, with such rapid advancements in the field and the integration of LLMs into voice assistance technology \cite{OpenAI2023, alexaLLM2023}, the capabilities of VAs are evolving too quickly for older adults to maintain a fixed mental model, which calls for a dynamic approach to continually update their understanding of these technologies.
To promote their continued use of VAs, it is crucial to develop clear, user-focused instructional protocols  for older adults \cite{koon2020perceptions}; including online tutorials and in-device instructions with real-time updates and corrective actions may effectively communicate VAs' capabilities and optimize their use among the older population.

VAs should also be designed to detect instances of user misconceptions, responding with tailored information that helps rectify such misunderstandings. Instead of defaulting to generic and uninformative responses such as \textit{``Sorry, I don't know that,''} the VA should provide context-specific explanations \cite{pradhan2020use, kim2021exploring}; for instance, if a user attempts to interact with a non-smart appliance, the VA should explain the prerequisites for smart home integration rather than merely stating, \textit{``Sorry, I couldn't find a connected light.''} This approach not only clarifies the VA's limitations but also educates the user about the necessary conditions for optimal use. Responses should be tailored to accommodate varying levels of technological familiarity, especially when considering use cases with older adults. 

\begin{adjustwidth}{0.3cm}{0pt}

    \textbf{Limitation 4:}
    \textit{A VA's inability to properly convey its capabilities results in continual inaccuracies in older adults' mental models---even amongst long-time users.}


    \noindent
    \textbf{Limitation 5:}
    \textit{Rapid advancements in voice assistance technology further exacerbate the gap between VAs' new capabilities and the relatively static mental models of older adults.}

   \noindent
   \hl{\textbf{Design Consideration 7:}
   \textit{VAs should employ context-aware responses with dynamic tutorials and real-time guidance in daily interactions to better align older adults' mental models with evolving voice assistance technology.}}

\end{adjustwidth}

\subsubsection{Lack of VA comprehension of users' natural conversation methods}

Our findings highlight a significant gap in Alexa's ability 
to comprehend natural conversational patterns, 
reflected by a high rate of intent recognition errors, which occur when the VA incorrectly responds to user queries ($32.3\%$, $n=204$ of $632$ as detailed in Section \ref{sec:error-types-resolution}). 
For instance, participants often preferred to use more conversational phrases such as ``Thank you'' 
instead of command-like terms such as ``Stop,'' likely perceiving the latter as pushy or rude; this could stem from the fact that older adults tend to favor polite interactions, both in their own communications to VAs \cite{pradhan2019phantom} and when receiving their responses \cite{hu2022polite}. 
Moreover, participants consistently reverted to natural phrases (\eg \textit{``Thank you, I have taken my medicine''}) instead of specific commands (\eg \textit{``Cancel alert''}) as illustrated in C4 in Table \ref{tab:alexa-dialogue-errors}, underscoring a desire for more intuitive VA interaction.
These intent recognition failures become particularly critical in high-stakes scenarios 
such as medication reminders; for example, P15 relied on Alexa for such reminders, but struggled to inform the VA when the task was completed (C4, Table \ref{tab:alexa-dialogue-errors}), leading to the confusion and frustration captured during one of her interactions: \textit{``I don't know what I am supposed to say to it to tell it that I have taken it.''} Despite the VA having the functionality to track completed reminders, it failed to understand the meaning behind P15's queries, such as ``\textit{Did I take my medicine today?}'' Older adults find the need to use specific keywords or commands annoying \cite{pradhan2020use, pradhan2018accessibility}. 
Compound errors (Section \ref{sec:errors-retry}) further escalate the friction between user and VA, as illustrated in C1 in Table \ref{tab:alexa-dialogue-errors}, wherein P2 kept reformulating his queries to get the information he wanted. 
Failure to recognize user intent can also be a reason for decreased usage over time; 
for instance, P1b discontinued her attempts to communicate with Alexa after repeated misunderstandings (C10, Table \ref{tab:alexa-dialogue-additional}).

Our data demonstrate that older adults adopt human dialogue as a metaphor for interaction with VAs \cite{cowan2017can, doyle2019mapping}, which can attributed to two factors: 1) speech interfaces are increasingly designed and marketed to emulate the nuances of human speech \cite{harris2004voice, gilmartin2017social} and 2) speech alone as interaction modality may inherently provide a false affordance for possible interactions---essentially tricking people into believing that natural human interactions are possible, while in reality, most VAs are only able to carry out structured human-machine interactions. These factors often lead to an overestimation of VAs' abilities to understand and conduct human-like conversations, resulting in communication breakdowns \cite{luger2016like, leahu2013categories, moore2016progress}. Furthermore, we observe that such misattribution of humanness to VA interactions is challenging to correct through trial and error, particularly among the aging population. This underscores the need for a more intentional design of voice-based interaction conventions---\ie \textit{natural human} interactions vs. \textit{structured human-machine} interactions---and for the more appropriate introduction and marketing of VAs.

Current commercial VAs require users to employ a \textit{structured human-machine} interaction style (\ie posing scripted commands and well-formulated queries), tying back to the previously discussed mismatch between users' mental models and VAs' capabilities in Section \ref{sec:discussion:mental-model}.
Older adults' tendencies to revert to a \textit{natural human} style of conversing highlights their expectation and preference for VAs to be more social, informal, and conversational, as opposed to younger \cite{oh2020differences} and middle-aged adults who have less trouble picking up on the nuances of structured human-machine interactions \cite{chin2024like}. Adapting VAs to understand and respond to more naturalistic conversational patterns---especially considering age-related difficulties for older adults (Section \ref{sec:discussion-age-factors}) and their slow adoption of new technologies \cite{vaportzis2017older, czaja2006factors}---is crucial for enhancing their user experience and ensuring effective communication. 

Therefore, intent recognition should be \textit{human-} and \textit{context-aware}, especially considering the expectations of older adults that VAs be more socially capable and the fact that such expectations 
differ both from other user groups \cite{oh2020differences} and within the older adult demographic itself \cite{harrington2022s}. For instance, leveraging people's verbal and vocal social cues to infer their implicit intent and creating user profiles can improve a VA's ability to accurately recognize user intent. Being agnostic to user state (\eg body language, audio-prosodic features, and other contextual information) results in a failure to appropriately respond or react to older adult's queries \cite{cuadra2022inclusion}. Moreover, as evidenced in our exploration of LLM-powered VAs (Section \ref{sec:results-chatGPT-breakdowns}) and as discussed in Sections \ref{sec:discussion-levraging-llm} and \ref{sec:discussion-speech-challenges}, utilizing LLMs' contextual capabilities (\eg retaining task information, conversation history, and user profiles) can improve error management---particularly by adapting to flexible user speech characteristics and preferences. As such, LLM integration presents a more effective approach to deciphering user intent \cite{mahmood2023llm}.

\begin{adjustwidth}{0.3cm}{0pt}

    \textbf{Limitation 6:}
    \textit{VAs' lack of comprehension of the naturalistic and diverse conversational styles practiced by older adults leads to intent recognition failures and subpar user experience.} 

    \noindent
    \hl{\textbf{Design Consideration 8:}
    \textit{Design VAs to be human- and context-aware so as to detect implicit intent through verbal and vocal cues.}}

  
\end{adjustwidth}  
  
\subsection{Limitations and Future Work}
While our four-week study provided insights into the behavior of older adults toward VAs, especially during interaction breakdowns, extended studies spanning several months could provide a deeper understanding in this area. Longer-duration interaction trends could additionally serve as a rich data source for training VAs to better recognize and recover from errors.
Currently, our analysis focuses qualitatively on queries and responses to assess their potential utility; however, a more detailed query analysis \cite{brewer2022empirical} would offer further insights into the nuances of older adults' interactions with VAs.
In this study, we investigated a generic application of LLM-powered VAs; future research should consider specific use cases, such as health care information retrieval or support tailored for older adults with disabilities and impairments. Additionally, our participant sample was relatively small and limited in scope, such as location (limited to one community center and local residents), cognitive ability (no individuals disclosed any memory problems), and ethnicity; future studies should aim for a larger and more diverse sample to encompass a wider range of experiences.



\section{Conclusion}

Traditional VAs often fail to effectively utilize immediate user reactions, users' explicit corrective actions, and conversation history in assisting users or recovering from errors. In our longitudinal study, we collected immediate user responses and reactions via a supplemental audio recorder, demonstrating that such social cues can often lead to increased error identification. Additionally, we found that VAs' current requirements for scripted commands and well-formed queries exacerbate conversational breakdowns with older adults, which may be caused by their inaccurate mental models of VAs' capabilities, their tendency to forget more effective query styles, and a preference for more naturalistic communication. We also explored integrating an LLM into an Alexa skill to enhance VAs' conversational capabilities with older adults; our findings provide insights and considerations for the design and research of nuanced interaction dynamics between older adults and futuristic conversational VAs, emphasizing the value of ``in-the-wild'' data collection to better comprehend such relationships.

\section*{Acknowledgements}
This work was supported by the National Science Foundation award \#1840088  and Malone Center for Engineering in Healthcare. We thank Jaimie Patterson for her feedback and assistance in this work.





\section*{CRediT author Statement}
\textbf{Amama Mahmood}: Conceptualization, Methodology, Software, Validation, Formal analysis, Investigation, Data curation, Writing - Original draft, Writing - Review \& editing, Visualization, Project Administration. 
\\
\textbf{Junxiang Wang}: Formal analysis, Writing - Original draft.
\\
\textbf{Chien-Ming Huang}: Conceptualization, Methodology, Resources, Writing - Original draft, Writing - Review \& editing, Visualization, Supervision, Funding acquisition. 

\newpage
\bibliographystyle{ACM-Reference-Format}
\bibliography{references}
\newpage
\section*{Appendix}
\input{appendix}

\end{document}

%% file: appendix.tex
\begin{appendix}
Supplementary materials are available at: \url{https://tinyurl.com/5av9wrdr}. 
\section{Data Analysis: Codebook}
\label{app:codebook}
The codebook used for labeling our interaction data is given in Table \ref{tab:codebook}. 
\begin{table}[b]
    \centering
    \caption{Our finalized codebook. Each one-turn query was labeled with codes listed below.}
    \label{tab:codebook}
    \begin{tabular}{p{3.5cm}p{12cm}}
\textbf{Code} &  \textbf{Definition} \\
\midrule
\midrule
Interaction type        & The purpose of the query. Some interaction types had subcodes (\eg ``questions'' were further labeled with question type).\\
Action      & The action associated with the interaction type, if applicable. For instance, music and radio had actions for ``play,'' ``pause,'' ``stop,'' etc. \\
Interaction        & The start and end of multi-turn interactions were marked to count the number of turns (queries) in each interaction.\\
Social response/reaction       & The immediate social response or reaction from the user.\\
Intended for Alexa      & Whether the user directed the query at Alexa or not.\\
Alexa handled      & Whether the query was transcribed by Alexa and Alexa acted or responded. \\
Alexa responded      & Whether Alexa responded verbally to the query or not. \\
Success              & Whether Alexa handled and responded to the query accurately.\\
Not handled         & The user query was captured accurately but Alexa did not respond. \\
Private conversation about Alexa  & The user engaged in a private conversation with themselves or others about Alexa during or immediately after the query. \\
Private conversation about same topic  & The user engaged in a private conversation with themselves or others about the same topic queried to Alexa during or immediately after the query. \\
Private conversation about how to use Alexa  & The user engaged in a private conversation with themselves or others about how to interact with Alexa during or immediately after the query. \\
Overlap & We coded separately whether the user talked over Alexa or vice versa. \\
Interruption & We coded separately whether the user interrupted Alexa or vice versa. \\
\hline 
\multicolumn{2}{c}{\textbf{For each erroneous interaction}} \\
\toprule
\textbf{Code} &  \textbf{Definition} \\
\midrule
\midrule
Error type & The type of error in the user query as determined by the coder from the transcribed logs.\\
Manifestation & Whether the coder could tell if an error occurred from the transcribed logs. \\
Identification & Whether the coder could tell if the user identified the error by their verbal response, reaction, or immediate action to fix the error. \\
Verbal/vocal reaction & The user's verbal or vocal reaction to the manifested error during or immediately after the query. \\
Retry & Whether the user tried to fix the error in the next immediate query. \\
Recovery strategy & The strategy employed by the user (if any) in the next immediate query to try and fix the error. \\
Resolution & Whether the error was resolved in the next immediate retry (if any) by the user. \\
Retries (n) & The number of retries taken by the user to fix a compounding error. The final query was marked with the total number of retries, signifying the conclusion of retries at that point for the original error. \\

\bottomrule
\end{tabular}

\end{table}

\begin{table}[tb]
\centering
\caption{Field study demographics: General. All participants were fluent in English. All were retired except P10a (employed full-time) and P11a (employed-part time). Note: eq. denotes ``equivalent diploma.''}
\label{tab:participants-full}
\begin{tabular}{lllllllll}

\textbf{P} &\textbf{Gender} & \textbf{Age} & \textbf{Ethnicity} & \textbf{Highest Degree} & \textbf{Profession} & \textbf{Disabilities/Aids} & \textbf{QOL Overall} \\
\midrule[1.5pt]
\multicolumn{8}{l}{\textbf{Community Center -- Assisted Living (by self)}} \\
\hline
2   & M & 66   & Prefer not to say & Bachelor's degree & Physician assistant & Wheelchair  &   Alright \\

3   & F & 82   & Caucasian & High school or eq. & Accountant clerk & Hearing aid &   Alright \\

4   & M & 79   & Caucasian & Master's degree & Actuary & Wheelchair &    Good \\
\hline
\multicolumn{8}{l}{\textbf{Community Center -- Independent Living (by self)}} \\
\hline
5   & F & 77   & Caucasian & Bachelor's degree & Social work &  & Good \\

6   & F & 81   & Caucasian & Master's degree & CIA & Cane/Rollator & Alright \\

7   & F & 74   & African American & Bachelor's degree & Administrator &  &    \\

14  & F & 73   & African American & Bachelor's degree & Senior claims & &  Good \\

15  & F & 73   & African American & Master's degree & Educator/Pastor &  &  Good \\
\hline
\multicolumn{8}{l}{\textbf{Homeowner (by self)}} \\
\hline
8   & F & 84   & Caucasian & Bachelor's degree & Writer/Editor & Walker & Alright \\

9   & M & 75   & Caucasian & Bachelor's degree & Case management & &   Very good \\

13  & M & 81   & Caucasian &  &  &  Vision issues  &    \\
\hline
\multicolumn{8}{l}{\textbf{Homeowner (couple, with spouse)}} \\
\hline
1a & M & 76   & Caucasian & Bachelor's degree & Business owner &  &    Good \\

1b & F & 75   & Caucasian & Nursing diploma & Nurse & &    \\
\cdashline{1-8}
10a & M & 68   & Caucasian & Bachelor's degree & Staff engineer &  &  Good \\

10b & F & 94   & Caucasian & Master's degree & Education & Walker/Scooter &   Bad \\
\cdashline{1-8}
11a & M & 72   & Caucasian & Master's degree & Urban planning & &   Very good \\

11b & F & 71   & Asian & Master's degree & Social work/Teacher & &    Very good \\
\cdashline{1-8}
12a & M & 75   & Caucasian & Bachelor's degree &  &  &  Very good \\

12b & F & 72   & Caucasian & Nursing diploma & Registered nurse &  &  Good \\
\cdashline{1-8}

\hline
\end{tabular}
\end{table}

\begin{table}[tb]
\centering
\caption{Field study demographics: Voice assistant technology use and familiarity.}
\label{tab:tech-use}
\begin{tabular}{l|l|p{12cm}}

\textbf{P} & \textbf{Placement}& \textbf{Technology use} \\
\midrule[1.5pt]
\multicolumn{3}{l}{\textbf{Community Center -- Assisted Living (by self)}} \\
\hline
2   & Studio & Familiar with ``the Alexa'' in the dining room of community center. Has Alexa app on iPad. \\
3   & Studio & Does not have smart phone. Never used VA before.\\
4   & Studio & Had Alexa a long time ago, but not anymore. \\
\hline
\multicolumn{3}{l}{\textbf{Community Center -- Independent Living (by self)}} \\
\hline
5   & Living room & Uses speech-to-text for messaging. Owns tablet. \\
6   & Living room & Uses TV remote via voice commands. \\
7   & Living room & Never used VA before. \\
14  & Living room & Uses Siri on iPhone to ask questions and for directions, locations, and restaurant ratings. Owns Apple Watch and iPad, as well. \\
15  & Living room & Never used VA before.  \\
\hline
\multicolumn{3}{l}{\textbf{Homeowner (by self)}} \\
\hline
8   & Bedroom & Does not have smart phone but owns iPad. Not familiar with VAs. \\
9   & Dining room & Owns a speaker compatible with Alexa, used mostly for music. Used occasionally for weather updates but primarily uses phone for this purpose. \\
13  & Dining room & Owns tablet. Not familiar with VAs. \\
\hline
\multicolumn{3}{l}{\textbf{Homeowner (couple, with spouse)}} \\
\hline
1 & Living room & Has Alexa smart speakers in living room, kitchen, bedroom, study, and at daughter's home next door. Used for communicating between rooms and timers while cooking, etc. \\
10 & Living room &   Not familiar with VAs. \\
11 & Kitchen & Uses Siri on iPhone for calls (especially when driving), sending messages, asking questions, and Google Maps.  \\
12  & Living room & Has Alexa smart speakers in kitchen (will move to laundry room) and bedroom. Used for communicating (drop-ins) between rooms, playing music, and listening to radio. \\
\hline
\end{tabular}
\end{table}

\end{appendix}